\documentclass[useAMS,usenatbib]{mn2e}

\usepackage{longtable}
\usepackage{pdflscape}
\usepackage{graphicx}
\usepackage{pdflscape}
\usepackage{color}
\usepackage{lscape,graphicx,pifont}

\newcommand{\Mdot}[2]{\mbox{${#1}\times10^{-{#2}}$\,M$_\odot$~yr$^{-1}$}}
\newcommand{\Msun}{\mbox{\,M$_\odot$}}
\newcommand{\Lsun}{\mbox{\,L$_\odot$}}
\newcommand{\vunit}{\mbox{\,km\,s$^{-1}$}}
\newcommand{\mic}{\mbox{$\,\mu$m}}
\newcommand{\pion}[2]{{#1}\,{\sc {#2}}}
\newcommand{\fion}[2]{[{#1}\,{\sc {#2}}]}

\newcommand{\ltsimeq}{\raisebox{-0.6ex}{$\,\stackrel
        {\raisebox{-.2ex}{$\textstyle <$}}{\sim}\,$}}
\newcommand{\gtsimeq}{\raisebox{-0.6ex}{$\,\stackrel
        {\raisebox{-.2ex}{$\textstyle >$}}{\sim}\,$}}

\newcommand{\va}{\mbox{V605~Aql}}

\newcommand{\fgs}{\mbox{FG~Sge}}

\newcommand{\sof}{\mbox{\it SOFIA}}
\newcommand{\SOFIA}{\mbox{\it Stratospheric Observatory for Infrared Astronomy}}
\newcommand{\sirtf}{\mbox{\it Spitzer}}
\newcommand{\spitzer}{\mbox{\it Spitzer Space Telescope}}

\newcommand{\chemone}{\raisebox{0.03cm}{$-$}} 
\newcommand{\chemthree}{\raisebox{0.03cm}{$\equiv$}} 

\title[Dust and molecules around Sakurai's Object]{The infrared view of dust 
and molecules around V4334~Sgr (Sakurai's Object): a 20-year retrospective} 
\author[A. Evans et al.]{A. Evans$^{1}$\thanks{E-mail: a.evans@keele.ac.uk},
R. D. Gehrz$^2$, 
C. E. Woodward$^2$, 
D. P. K. Banerjee$^{3}$, 
T. R. Geballe$^4$, \newauthor
G. C. Clayton$^5$, 
P. J. Sarre$^{6}$, 
S. Starrfield$^{7}$,
K. Hinkle$^{8}$, 
R. R. Joyce$^{8}$, \newauthor
Foteini Lykou$^{9,10}$, 
L. A. Helton$^{2,11}$, 
S. P. S. Eyres$^{12,13}$,
H. Worters$^{14}$, \newauthor
E. J. Montiel$^{11,15}$, 
T. Liimets$^{16,17}$, 
A. Zijlstra$^{18,9}$,  
M. Richter$^{15}$,
J. Krautter$^{19}$
 \\ \\ %
$^{1}$Astrophysics Group, Keele University, Keele, Staffordshire, ST5 5BG, UK\\
$^{2}$Minnesota Institute for Astrophysics, School of Physics \& Astronomy,
116 Church Street SE, University of Minnesota, \\ \mbox{~~~} Minneapolis, MN 55455, USA\\ 
$^{3}$Physical Research Laboratory, Ahmedabad 380009, India\\ 
$^{4}$Gemini Observatory, 670 N. Aohoku Place, Hilo, HI, 96720,
USA\\ 
$^5$Department of Physics and Astronomy, Louisiana State University, Baton Rouge, LA 70803, USA \\ 
$^{6}$School of Chemistry, The University of Nottingham, University Park, Nottingham, NG7 2RD, UK \\ 
$^{7}$School of Earth and Space Exploration, Arizona State University, Box 871404, Tempe, AZ 85287-1404, USA\\ 
$^{8}$NSF's National Optical-Infrared Astronomy Research Laboratory, PO Box 26732, 
Tucson, AZ 85726, USA \\ 
$^{9}$Department of Physics, The University of Hong Kong, C.Y.M. Physics Building, Pokfulam Road, Hong Kong\\ 
$^{10}$Laboratory for Astrophysics, The University of Hong Kong, 100 Cyberport Road, Cyberport, Hong Kong \\ 
$^{11}$USRA-SOFIA Science Center, NASA Ames Research Center, Moffett Field, CA 94035, USA\\ 
$^{12}$Faculty of Computing, Engineering \& Science, University of South Wales, Pontypridd, CF37 1DL, UK\\ 
$^{13}$Jeremiah Horrocks Institute, University of Central Lancashire, Preston PR1 2HE, UK\\ 
$^{14}$South African Astronomical Observatory, PO Box 9, Observatory, Cape Town 7935, South Africa \\ 
$^{15}$Department of Physics, University of California Davis, 1 Shields Ave, Davis, CA 95616, USA \\ 
$^{16}$Tartu Observatory, University of Tartu, Observatooriumi 1, 61602 T\~oravere, Estonia \\ 
$^{17}$Astronomick\'y \'ustav, Akademie v\v{e}d \v{C}esk\'e republiky, v.v.i., Fri\v{c}ova 298, 251\,65 Ond\v{r}ejov,  Czech Republic\\ 
$^{18}$Jodrell Bank Centre for Astrophysics, School of Physics and Astronomy, University of Manchester, Manchester, M13 9PL, UK \\ 
$^{19}$Landessternwarte-Zentrum f\"ur Astronomie der Universit\"at, K\"onigstuhl, D-69117 Heidelberg, Germany\\ 
}

\begin{document}

\date{Version of 2019-06-06}

\pagerange{\pageref{firstpage}--\pageref{lastpage}} \pubyear{2019}

\maketitle

\label{firstpage}

\begin{abstract} 
We present an analysis of the evolution of circumstellar dust and 
molecules in the environment of the very late thermal pulse object V4334~Sgr
(Sakurai's Object) over a $\sim20$-year period, drawing on  
ground-, airborne- and space-based infrared photometry and spectroscopy. 
The dust emission, which started in 1997,  resembles a blackbody
that cooled from $\sim1,200$~K in 1998 
August to $\sim∼180$~K in 2016 July. The dust mass, assuming 
amorphous carbon, was $\sim5\times10^{-10}$\Msun\ in 1998 August, and we
estimate that the total dust mass was $\sim2\times10^{-5}$\Msun\ 
by $\sim2016$. The appearance of a near infrared excess in 2008 suggests 
a new episode of (or renewed) mass loss began then.
We infer lower limits on the bolometric luminosity of the 
embedded star from that of the dust shell, which rose to $\sim16,000$\Lsun\ 
before declining to $\sim3,000$\Lsun. There is evidence for weak 6--7\mic\ 
absorption, which we attribute to hydrogenated amorphous carbon formed in
material ejected by Sakurai's Object during a 
mass ejection phase that preceded the 1997 event. 
We detect small hydrocarbon and other molecules in the spectra, and trace the 
column densities in hydrogen cyanide (HCN) and acetylene (C$_2$H$_2$).
We use the former to determine the $^{12}$C/$^{13}$C ratio to be
$6.4\pm0.7$, 14 times smaller than the Solar System value.
\end{abstract}


\section{Introduction}

It is well-known that the fate of a star when it evolves away from the Main
Sequence (MS) depends on its mass. The ``text-book'' scenario \citep{textbook} for
the post-MS evolution of low to inter\-mediate mass stars is that, 
following the helium core flash, burnout of He occurs in the core on the horizontal
branch. After evolution up the Asymptotic Giant Branch (AGB), the star sheds its
outer envelope, which is illuminated as a planetary nebula (PN) by the still-hot
stellar core. In time the PN disperses and its nucleus becomes a white dwarf (WD).

However, in as many as 10--20\% of cases \citep{blocker,lawlor} the star,
even after it has taken the left turn at the ``knee'' beyond the post-AGB 
phase on the Hertzsprung-Russell (HR) diagram and is evolving downwards 
towards the WD region, re-ignites a residual helium shell in a Very Late 
Thermal Pulse (VLTP). It then retraces a large part of its evolutionary track 
\citep{iben,herwig,lawlor} and becomes a ``Born-Again Giant'' (BAG).
A number of examples of the VLTP phenomenon are now known, including Sakurai's
Object (V4334~Sgr, hereafter ``Sakurai'') and V605~Aql.
FG~Sge has also been regarded as a BAG \citep{gehrz-fgsge},
although it is also regarded as a Late Thermal Pulse \citep{JS}.
All these stars are hydrogen-deficient, carbon-rich,
have extensive dust shells, and each lies at the centre of an old PN.

The final evolution from pre-WD to the BAG phase is predicted to take of the order 
of a few centuries, thus representing a very rapid (and hence seldom seen) 
phase of stellar evolution. However the rate at which Sakurai has evolved has
radically changed our views about the post-MS evolution of low-to-intermediate
mass (1--8\Msun) stars, prompting a major rethinking of the underlying
astrophysics \citep{herwig,herwig05,lawlor}.

Here we present a series of infrared (IR) observations of Sakurai, obtained with 
ground-based, airborne and space-based observatories over 
the 20-year period since it became dust embedded in late 1998.
We include already-published and unpublished
data, and data acquired recently from the NASA \SOFIA\ 
\citep[\sof;][]{sofia,sofia2}.
An overview of the early (1996--2001) IR evolution -- including the molecular 
and dust components -- was given by \cite{geballe}.

\section{Sakurai's Object (V4334~Sgr): A Brief History of its Evolution}
\label{history}
V4334~Sgr was discovered by the amateur astronomer Yukio Sakurai in 1996 
\citep*[see][]{nakano}. Originally reported as a nova, it was at first 
spectroscopically similar to an F~supergiant, possibly with a hot dust shell
\citep{duerbeck96}. 

It is now known to be a low-mass ($\sim0.6$\Msun) post-AGB star that is 
retracing its post-AGB evolution along the HR diagram following a VLTP 
\citep{herwig}. This interpretation was strengthened by the discovery that 
Sakurai lies at the centre of a faint planetary nebula (PN) $40''$ in 
diameter \citep[e.g.][]{kerber,eyres-rad,pollacco}. \citeauthor{pollacco}
measured the expansion velocity of the PN shell to be $32\pm6$\vunit\ which,
for a distance of 3.8~kpc (see below), gives a ballistic expansion age of 
$\sim11,300$~years. This PN age probably rules out a progenitor having mass 
$\ltsimeq1.25$\Msun, as the progenitor would not have reached the WD cooling 
track in the age of the Galaxy. Furthermore, the location of the central star of the PN 
on the HR diagram in 1995 is incompatible with its being a pre-WD \citep[see][]{herwig}.

In mid-1996, the spectrum of Sakurai rapidly evolved to that of a C-rich, 
H-poor star, with elemental
abundances akin to those found in the R~Coronae Borealis (RCB) stars
\citep{asplund99}. Subsequently, around 1997 April, 
Sakurai started to produce a carbon dust shell that obscured the central star
\citep[see Figure~1 of][for a visual light curve to mid-1999]{duerbeck02}. 
As the visual light faded, photospheric CO was detectable in the IR
until late 1998 but thereafter, even the IR spectrum was
completely devoid of photospheric features.

By 1999 April the photosphere had completely disappeared from view, 
even at IR wavelengths, and as of early 2020, remains obscured. A 
large mass-loss rate ($\sim10^{-5}$\,\Msun~yr$^{-1}$ over the period 
1999 May -- 2001 September) is implied by 1--5\mic\ spectroscopy 
\citep{tyne-dusty}. \cite{evans-jcmt1,evans-sak} detected the dust around 
Sakurai at 450\mic\ and 850\mic\ in 2001 August, and found that the flux
density at these wavelengths was still rising in 2003, indicating that the
mass-loss was being maintained; the spectral 
energy distribution (SED) resembled a black-body at 360~K.

High spatial resolution observations using MIDI/VLTI 
\citep{chesneau} resolved Sakurai's dust disc, which has an asymmetry that is 
aligned with the PN reported by \cite{pollacco}; the inclination of the dust 
disc relative to the plane of the sky is $\sim75^\circ$. 
The mass of dust in the disc was determined to be 
$\sim6\times 10^{-5}$\Msun, although this value is somewhat model-dependent.

The optical spectrum has shown significant changes since $\sim2008$
\citep[e.g][]{vanhoof15,steene}, when the strength of many emission lines started
to increase, most notably \fion{O}{ii} $\lambda7320/7330$\AA. In addition
\citeauthor{steene}  identify the emergence of electronic transitions of CN
at this time, with a likely origin in the dust disc.

On the basis of stellar evolutionary models, \cite{lawlor} suggested 
that the rapid evolution (on a timescale $\sim10$~yr) of
Sakurai and \va\ --  compared with the longer timescale associated with 
the evolution of  \fgs\ -- might indicate that the 
latter is making its second visit to the AGB, whereas Sakurai and \va\ are visiting 
for the first time. They predicted that the stellar component of Sakurai would rapidly 
increase in effective temperature, and will eventually resemble \va, whereas the 
latter will return to the AGB over the next several decades.
That Sakurai and \va\ are evolving along similar evolutionary tracks, with 
\va\ somewhat more advanced by virtue of its having undergone a thermal pulse 
in 1919, was also concluded by \cite{clayton,clayton2}. 
Indeed \citeauthor{clayton} suggested that the behaviour of 
\va\ could serve as a template for the future behaviour of Sakurai.

\cite{blocker} argued that the stellar evolution pathway taken by
Sakurai is a fate that awaits 20\% of low-mass stars; thus we are looking at
a possible fate of the Sun. 
\cite{herwig-i} and others have studied details of the
VLTP, and find that mixing of protons into the He-burning shell 
can trigger i-process nucleosynthesis, producing the anomalous abundances of 
species such as Sr, Y, and Zr seen in Sakurai \citep{asplund97}.
Furthermore, understanding objects 
that have undergone VLTPs has a much wider significance: there is 
evidence that carbon stars produce dust in low metallicity environments, 
with implications for dust production in (for example) globular clusters and
dwarf galaxies \citep{sloane,boyer-15a,boyer-15b,boyer-17,goldman}.

\section{Dust \& Molecules}

\subsection{The dust}

\subsubsection{The start of the dust phase}
\label{start}
By late-1998, the central star of Sakurai was completely obscured by 
an optically thick dust shell.  \cite{evans-sak-sp} reviewed the 
expansion of the dust shell by estimating its
``blackbody angular diameter'' (see Appendix~A). They found that the 
expansion was linear in time, with an extrapolated start on 1997 November 10 
(MJD\footnote{Throughout we use Modified Julian 
Date (MJD) to identify the time of the observations. MJD is related to Julian 
Date (JD) by $\mbox{MJD} = \mbox{JD} - 2400000.5$} 50762); 
the expansion rate was $0.059$~mas~d$^{-1}$. \cite{hinkle} determined a similar 
expansion rate ($0.055\pm0.008$~mas~d$^{-1}$)  over a somewhat 
longer time-base, also with a start in 1997 November.
In either case the extrapolation to zero diameter coincided with the start of 
dust production sometime between 1997 Nov 7 
(MJD~50759)
and 1998 February 5 \citep[MJD 50849;][]{duerbeck02}.
However, as noted in Section~\ref{history}, the IR spectrum shows that 
dust had actually begun making a contribution to the IR emission as early as 1997 April 
\citep{geballe,pavlenko02}.

Since 1998, the 1-5\mic\ 
continuum spectral energy distribution (SED) has crudely resembled a
featureless blackbody, commensurate with that emitted by an
optically thick carbon dust shell. We will confine our analysis of the dust
shell to the period after 1997 November.

\subsubsection{The nature of the dust}
\label{nature}
Analysis of the post-1998 IR observations shows that the dust is 
carbon-based, primarily in amorphous form \citep{,eyres-ir,tyne-dusty,evans-sak-sp}.
The 1-5\mic\ spectra during the period 1999--2001 were consistent with mass-loss 
rate ({\em in the form of dust only}) that increased steadily from 
$\sim\Mdot{5}{8}$ to 
$\sim\Mdot{1.8}{7}$ \citep[see][for details]{tyne-dusty}, where we have scaled 
as per the {\sc dusty} code \citep{dusty} for the distance $D$ and
the density of the grain material (see Section~\ref{Mdusts}
for the value assumed here).
A crude integration of the mass-loss rate (in dust only) over the period 
1999--2001 from \cite{tyne-dusty}
yields the accumulated mass of dust $M_{\rm dust}$ ejected to the end of 2001 to 
be $\simeq1.0\times10^{-5}$\Msun. 

The dust mass over the period 2001 August to 2003 June was determined
by \cite{evans-sak} using 850\mic\ photometry. These authors found that the 
dust mass increased from $2\times10^{-6}$\Msun\ to $1.4\times10^{-5}$\Msun\
over this period (again these values are scaled to the distance assumed here).

The maximum grain radius (in a grain size ($a$) distribution 
$n(a)\,da \propto a^{-q}\,da$, 
with $q\simeq3$) increased by a factor $\sim3$ over the period 1999 May to 2001 
September \citep{tyne-dusty}.

\begin{figure}
\setlength{\unitlength}{1mm}
\begin{center}
\leavevmode
  \includegraphics[width=9cm,height=7.5cm]{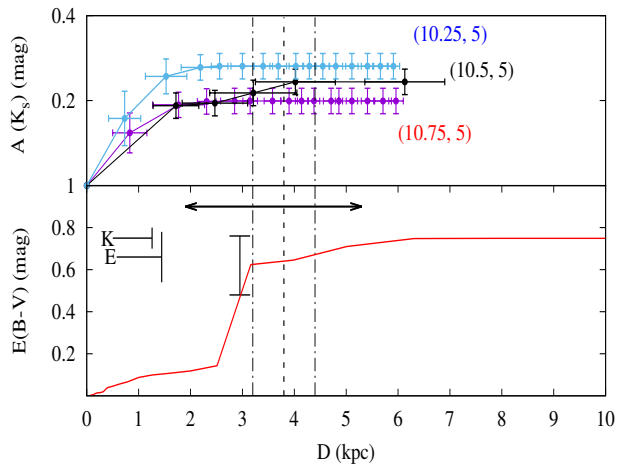}
  \caption[]{Dependence of reddening $E(B-V)$ \citep[from][red curve, bottom
  panel]{green}, 
  and of $K_s$ band extinction \citep[from][top panel]{marshall}, on distance. 
  The $K_s$ band extinction is for three lines of sight close to Sakurai
  ($l=10.48^\circ$, $b=+04.41^\circ$); the numbers in
  brackets give $(l,b)$. Horizontal double-headed arrow gives range of
  possible distance from \cite{eyres-rad}.
  Lines labelled ``K'' and ``E'' give reddening from \cite{kimeswenger} 
  (the weighted mean of several values), and 
  \cite{evans-red} (based on strength of the interstellar silicate
  absorption feature). 
  Error bar at distance $\simeq2.6$~kpc gives reddening as detemined here.  
  Broken vertical lines depict the adopted distance 
  (3.8~kpc) and associated uncertainties. See Section~\ref{RD} for details. \label{AKs}}
\end{center}
\end{figure}

\subsection{The molecules}

IR observations \citep{eyres-ir} revealed the presence of C$_2$, CN and 
CO through their electronic \citep[C$_2$;][]{BR} and vibrational transitions, 
including overtone absorption spectra of $^{12}$CO 
and $^{13}$CO. The presence of the Ballik-Ramsay bands
mirrored the appearance of the C$_2$ Swan bands \citep{swan} in the optical \citep{asplund97}. 
The C$_2$ bands suggested a $^{12}$C/$^{13}$C ratio in the range 1.5--5. 
High resolution ($R\simeq30,000$) spectroscopy  \citep{pavlenko04} resolved 
the first overtone $^{12}$CO and $^{13}$CO bands, to which
\citeauthor{pavlenko04} fitted a synthetic hydrogen-deficient model 
atmosphere to determine a $^{12}$C$/^{13}$C ratio of $4\pm1$. Both 
values are substantially less than the Solar System 
value \citep*[$^{12}$\mbox{C}/$^{13}\mbox{C} = 89$;][]{lodders}.
\cite{worters} used the IR CO fundamental bands
-- seen in absorption against the dust emission -- to
derive a ratio similar to that of \cite{pavlenko04}.
\cite{evans-sak-sp}, using mid-IR HCN lines, also found that 
the $^{12}$C/$^{13}$C ratio is $\sim4$. 

\onecolumn

\setlongtables
\LTcapwidth=8.4in
\small

\begin{landscape}
 
\begin{table*}
\caption{Evolution of Sakurai's dust as seen from UKIRT, 2MASS, {\it WISE}, \spitzer\ 
and \sof. Column headed ``Refs'' lists papers containing IR data on Sakurai already published. 
Observations in the $K_s$ band are listed separately 
(Table~\ref{vvv}). See text for details. \label{spitzer}}
\begin{tabular}{ccclccccccc}
Facility &  Date       &Ref$^*$& \multicolumn{1}{l}{MJD}       
                                    & Wavelength coverage & $R^{**}$    & $T_{\rm dust}$ & $[\lambda f_\lambda]_{\rm max}$ & $L_{\rm dust}^\dag$ & $M_{\rm dust}^\dag$\\
         & YYYY-MM-DD  &&            &        ($\mu$m)   &         & (K)  &   ($10^{-12}$ W m$^{-2}$) & (\Lsun)  & (\Msun) \\ \hline 
 UKIRT  & 1998-08-18  & 1 &   51043   &  1.2--2.3   &       & $1,210\pm3$ & $1.29\pm0.01$ & $791\pm177$ & $5.10[\pm1.14]\times10^{-10}$\\
 $''$   & 1999-04-21  & 2 &   51289   &  1.0--2.5   &  690--950     & $840.6\pm0.6$  & $8.16\pm0.05$    & $5,004\pm1,118$ & $1.82[\pm0.41]\times10^{-8}$\\
2MASS    & 1999-04-29  &&  51297      &  $J\!H\!K_s^\ddag$ & $\sim6$  & $950\pm37$ & $6.57\pm2.48$ & $4,029\pm1,767$ & $8.21[\pm3.91]\times10^{-9}$ \\
UKIRT    & 1999-05-03  & 3 &  51301   &  1--5        &  1,300--4,800    & $723.5\pm0.5$ & $14.17\pm0.12$ & $8,689\pm1,942$ & $6.46[\pm1.44]\times10^{-8}$\\
$''$     & 1999-06-08  & 3 & 51337    &  1--5        &  200--2,000   & $717.1\pm1.8$ & $14.81\pm0.23$ & $9,081\pm2,033$ & $7.04[\pm1.58]\times10^{-8}$\\
$''$    & 1999-06-14  & 3 & 51343     & 1--5         & 200--2,000 & $717.2\pm1.1$  & $14.89\pm0.04$ & $9,130\pm2,039$ & $7.08[\pm1.58]\times10^{-8}$\\
$''$    & 1999-09-06  & 3 & 51427     &  1--5        & 200--2,000 & $660.9\pm0.7$  & $15.66\pm0.11$ & $9,602\pm2,145$ & $1.10[\pm0.25]\times10^{-7}$\\
$ ''$   & 2002-04-14  & 3 & 52378     & 1.4--5       & 200--2,000 & $432.4\pm0.3$  & $26.57\pm0.13$ & $16,292\pm3,639$ & $1.40[\pm0.31]\times10^{-6}$\\
 $''$  & 2003-09-08  & 3 &  52890     &  1--5        & 200--2,000 & $359.6\pm0.2$ & $12.40\pm0.05$&  $7,604\pm1,698$ & $1.57[\pm0.35]\times10^{-6}$\\             
\sirtf   & 2005-04-15  & 4 &  53475.3 & 5.2--36      &  60--600 & $283.5\pm0.2$  & $13.17\pm0.05$ & $8,076\pm1,804$ & $5.16[\pm1.15]\times10^{-6}$\\
 $''$  & 2007-05-04  &&  54224.9      &  5.2--33     & $''$ & $225.5\pm0.1$ & $9.58\pm0.03$ &  $5,874\pm1,312$ & $1.11[\pm0.25]\times10^{-5}$ \\
 $''$  & 2007-10-15  && 54388         & 5.2--33      & $''$ & $217.0\pm0.3$  & $9.05\pm0.05$ &  $5,549\pm1,240$  &  $1.26[\pm0.28]\times10^{-5}$ \\
  $''$ & 2008-04-08  &&   54564       &  5.2--33.2   & $''$       & $205.8\pm0.2$& $8.49\pm0.03$ & $5,206\pm1,163$ & $1.53[\pm0.34]\times10^{-5}$\\
 $''$  & 2008-10-18  &&  54757.1      &  5.2--35     & $''$        & $207.3\pm0.2$& $8.52\pm0.03$&  $5,224\pm1,167$ & $1.48[\pm0.33]\times10^{-5}$ \\
{\it WISE}     & 2010-03-19  &&  55274.4   &   W1,W2,W3,W4$^{\S,\S\S}$   &  1--3  &--- & --- & --- & --- \\
\sof   & 2014-03-25  &&  56741.5      &  8.6--36.6   & 110--160     &$181.7\pm0.6$ & $5.47\pm0.07$& $3,354\pm750$ & $1.78[\pm0.40]\times10^{-5}$\\
 $''$  & 2016-07-11  &&  57580.3      &   5.0--31    & 110--180    & $176.9\pm0.5$& $4.95\pm0.06$ & $3,035\pm679$ & $1.83[\pm0.41]\times10^{-5}$ \\ \hline
 \multicolumn{7}{l}{$^{*}$1: \cite{geballe}; 2: \cite{tyne1}; 3: \cite{tyne-dusty}; 4: \cite{evans-sak-sp}.} & & & \\
 \multicolumn{4}{l}{$^{**}$Spectral resolution~$R=\lambda/\Delta\lambda$.} & & & \\
 \multicolumn{4}{l}{$^\dag$For $D=3.8$~kpc.} & & & \\
 \multicolumn{4}{l}{$^\ddag$2MASS pass bands.} & & & \\
  \multicolumn{4}{l}{$^\S${\it WISE} pass bands.} & & & \\
 \multicolumn{4}{l}{$^{\S\S}$Data  saturated in W3 and W4; see text.} & & & \\
       \end{tabular}
\end{table*}

\end{landscape}

\twocolumn

\cite{tafoya} detected the $J=4\rightarrow3$ transitions of H$^{12}$CN and 
H$^{13}$CN, and the $J=1\rightarrow0$ transition of H$^{12}$CN, and found
that the expansion velocities were in excess of  $100$\vunit.
They too found that the $^{12}$C/$^{13}$C ratio is extremely low.  
The consistently low $^{12}$C/$^{13}$C ratio in Sakurai -- determined from optical, 
IR and sub-mm observations -- is in line with that expected 
\citep[$\sim3.5$;][]{asplund99} 
from second stage equilibrium CNO burning following He-burning during the VLTP, 
and therefore with Sakurai being an object that has undergone a VLTP.

\section{Reddening and Distance}
\label{RD}
While the assumed interstellar reddening to Sakurai will have minimal bearing on 
most of the data we discuss here, there will
be a discernible effect at the shortest (${\sim}J\!H\!K$) wavelengths.

\subsection{Distance}

A search of the GAIA DR2 archive \citep{gaia} reveals no parallax values for
Sakurai. Various distance determinations from the literature
are summarised by \cite{HJ}, 
in which nearly all estimates lie within the 
range proposed by \cite{eyres-rad}, namely 1.9--5.3~kpc.
As pointed out by \cite{vanhoof7}, the line-of-sight to
Sakurai reaches the scale height of the old disc at $D=4$~kpc, 
making this an upper limit (albeit a weak one) on the distance.

We adopt here the ``statistical'' distance deduced by 
\cite{eyres-rad}, namely $3.8\pm0.6$~kpc. We note that the 
expansion of the black body angular diameter of the dust (see 
Section~\ref{start}), combined with this distance, leads to 
an expansion velocity for the dust-forming material of $\sim375$\vunit; this is comparable with that suggested by the blue wing of
the \pion{He}{i} 1.0833\mic\ line in 1997 July \citep{geballe}.

\subsection{Reddening}

Methods of determining the interstellar reddening, $E(B-V)$, to Sakurai have 
relied mainly on measuring the Balmer decrement. An early determination 
using this method
was made by \cite{duerbeck96}, who found $E(B-V)=0.54$~mag.
However more recent considerations have deduced larger reddening.
\cite{kimeswenger} gives a weighted mean, based on several independent 
determinations up to 2002, of $0.75\pm0.05$~mag. \cite{evans-red} reported 
a detection of the 9.7\mic\ silicate feature in absorption in Sakurai.
They argued that the feature actually arises in the interstellar medium, 
and therefore provides a direct measure
of the reddening by interstellar material, $E(B-V)=0.66 \pm 0.12$~mag.
\cite{vanhoof7} found $E(B-V)=0.86$~mag from the Balmer decrement, assuming 
that ``Case~B'' applies.

\cite{marshall} have used the 
Two Micron All Sky Survey \citep[2MASS;][]{2mass} catalogue to map out 
the Galactic extinction in the $K_s$ (2.159\mic) band;
in the direction of Sakurai their data indicate $A_{K_s}\simeq0.20$ 
to 0.24~mag for the range of distances suggested
by \cite{eyres-rad}, as depicted in Fig.~\ref{AKs}. A weighted mean of 
the $A_{{K_s}}$ values in \citeauthor{marshall}, over the \citeauthor{eyres-rad} 
distance range, gives $A_{{K_s}}\simeq0.22\pm0.03$~mag. This value of extinction
is consistent with the extinction analysis by \cite{gonzalez},
who found $E(J-K)=0.312\pm0.115$~mag and $A_K=0.216$~mag towards Sakurai. 
The $A_{{K_s}}$  value of $0.22\pm0.03$~mag corresponds
to $E(B-V)=0.61\pm0.07$~mag, close to the value derived by \cite{evans-red}.

These considerations suggest that determinations of the redd\-ening 
using the Balmer decrement and Case~B may overestimate $E(B-V)$.
On the other hand determining the reddening using only the agents of interstellar
extinction points to a consistent $E(B-V)=0.61\pm0.07$~mag ($A_{K_{s}}$ extinction)
and $E(B-V)=0.66\pm0.12$~mag (9.7\mic\ feature).
For the purposes of this paper we use a weighted mean of the reddening as deduced 
from the interstellar extinction only, namely $E(B-V)=0.62\pm0.06$~mag.

\begin{figure*}
\setlength{\unitlength}{1mm}
\begin{center}
  \leavevmode
      \includegraphics[width=10cm]{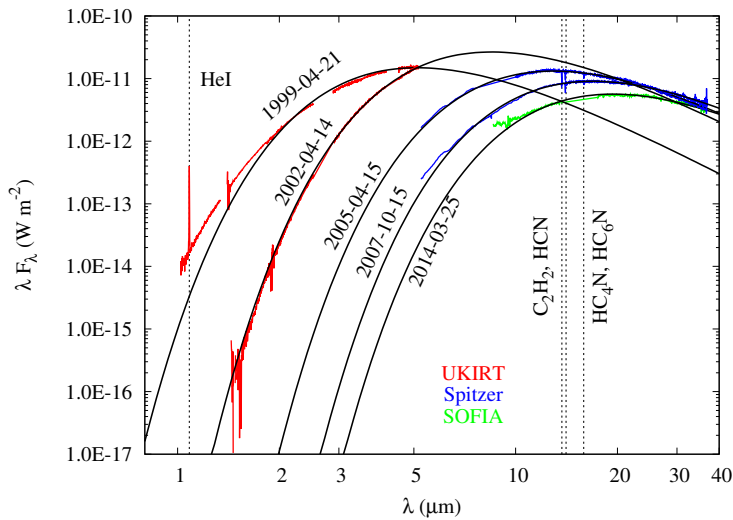}
    \caption[]{Selected snapshots of the evolution of the SED of Sakurai's dust shell, 
    obtained on the dates indicated and with the facilities shown, from 1999 to 2014. 
    Note the poor atmospheric transmission
    in the UKIRT data at wavelengths $\sim1.4$\mic\ and $\sim1.8$\mic.
    A warm dust contribution, not shown, is present in the near-IR in 2014-03-25
    (see Section~\ref{fresh}).
    Curves are labelled with the dates of observation; see 
    Table~\ref{spitzer} for further details.
    Wavelengths of the \pion{He}{i} 1.083\mic\ line, and the hydrocarbon lines, are 
    indicated. \label{sak_all}}
\end{center}
\end{figure*}

\section{The Data}

We have compiled a series of IR observations -- both photometric and 
spectroscopic -- that show the dust emission by Sakurai, covering the 
period 1996 April to 2016 July, providing a 20 year retrospective of 
IR observations of the dust evolution of a VLTP. We briefly describe the 
observations here. An observing log is given in Table~\ref{spitzer} 
and a sample of the spectroscopic data,
together with blackbody fits (discussed in Section~\ref{dustemission} below), 
is shown in Fig.~\ref{sak_all}.

\subsection{UKIRT}

Sakurai was observed with the United Kingdom Infrared Telescope (UKIRT) from 1996 to
2000. Full details of the spectrometers used, data acquisition and reduction,
together with the data themselves, are given in \cite{eyres-ir}, 
\cite{pavlenko04}, \cite{tyne-dusty} and references therein.

\subsection{$K_s$ band photometry}
\label{Ks}
\begin{figure}
\setlength{\unitlength}{1mm}
\begin{center}
  \leavevmode
  \includegraphics[width=8cm]{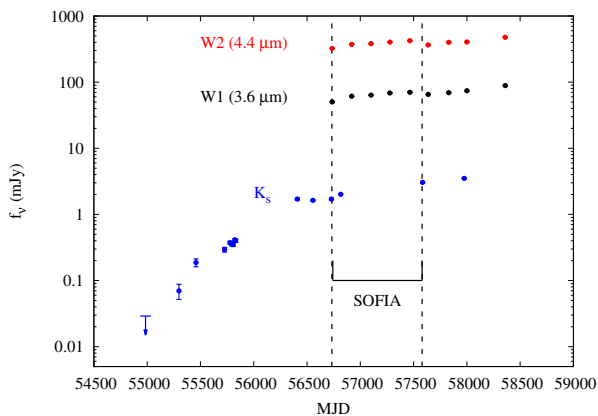}
    \caption[]{$K_s$-band (blue), {\it NeoWISE} bands W1 (black) and W2 (red)
    light curves of Sakurai. Times of the \sof\ observations discussed in 
    Section~\ref{sofia} are indicated; see Section~\ref{fresh} for discussion of data highlighted by the dotted vertical line. \label{K_lc}}
\end{center}
\end{figure}

Archival data on Sakurai have been retrieved from the European Southern Observatory's 
{\it VISTA Variables in the V\'ia L\'actea} survey \citep{saito}
and are listed in Table~\ref{vvv}, which also includes data from \citeauthor{HJ2} 
({\citeyear{HJ2}; see that paper for instrumental details)
and \cite{HJ3}. The observations on MJD 55457.25/55458.21 were combined into a 
single image, which shows both a bright point source and ejecta. Aperture 
photometry was obtained on both images and the
the fluxes were combined to get the measured magnitude.
The data are plotted in Fig.~\ref{K_lc}.

\begin{table}
\caption{$K_s$-band observations of Sakurai. 
See references below for full details.\label{vvv}}
 \begin{tabular}{lccc} 
   MJD & $K_s$ (mag) & Ref & Comment \\ \hline
54984.79   & $>18.4$    &  2 & $3\sigma$ limit \\ 
55298.27   & $17.45\pm0.28$  & 1 & \\
55457.25   &                 &   & Average of \\
55458.21   & $16.38\pm0.15$  & 2 & two nights\\
55726.31   & $15.89\pm0.09$  & 1 & \\  
55777.19   & $15.63\pm0.06$  & 1 & \\  
55791.09   & $15.68\pm0.06$  & 1 & \\  
55807.08   & $15.70\pm0.07$  & 1 & \\  
55823.06   & $15.52\pm0.05$  & 1 & \\  
55831.01   & $15.56\pm0.06$  & 1 & \\  
56408.53   & $13.98\pm0.04$  &  2 & \\
56555.10   & $14.03\pm0.02$  &  2 & \\
56729.52   & $13.98\pm0.04$  &  2 & \\
56816.33   & $13.80\pm0.01$  &  2 & \\
57584.91   & $13.35\pm0.02$  &  3 & \\ 
57975.07   & $13.20\pm0.02$  & 1 & \\  
 &&  \\ \hline
 \multicolumn{4}{l}{1. Data from VISTA InfraRed CAMera on 
 the 4~m Visible}\\
 \multicolumn{4}{l}{and Infrared Survey Telescope for Astronomy (VISTA).}\\
 \multicolumn{4}{l}{See \cite{saito}.} \\
 \multicolumn{4}{l}{2. Data obtained with the High-Resolution  Infrared Camera}\\ 
 \multicolumn{4}{l}{on the Kitt Peak National Observatory (KPNO)  3.5m telescope,}\\ 
 \multicolumn{4}{l}{the Extremely Wide-Field Infrared Imager on  the KPNO 4-m}\\ 
 \multicolumn{4}{l}{telescope and with the Near InfraRed Imager and  spectrograph}\\ 
 \multicolumn{4}{l}{(NIRI) on Gemini~North telescope. See \cite{HJ2}.} \\
 \multicolumn{4}{l}{3. Data obtained with NIRI. See \cite*{HJ3}.}\\
  \end{tabular}
\end{table}

\begin{figure}
\setlength{\unitlength}{1mm}
\begin{center}
  \leavevmode
  \includegraphics[width=8cm]{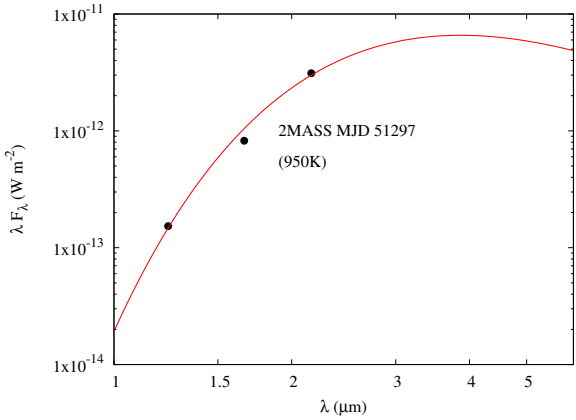}
    \caption[]{SED of Sakurai from 
    2MASS survey photometry; red curve is blackbody fit.\label{2mass-fig}}
\end{center}
\end{figure}

\subsection{2MASS}
Sakurai is included in the 2MASS survey, with 
fluxes of $38.08\pm0.66$~mJy, $341.27\pm9.24$~mJy and $1884\pm19$~mJy in the $J$ (1.235\mic), 
$H$ (1.662\mic) and $K_s$ (2.159\mic) bands respectively.
The data are for the epoch 1999 April 29 (MJD~51297) and are shown in 
Fig.~\ref{2mass-fig}.

\subsection{\it WISE}
\label{wise}

Sakurai was observed in eleven scans during the Wide-field Infrared Survey 
Explorer \citep[{\it WISE};][]{wise} survey
during a $\sim1$ day interval centered around MJD~55274.4. 
The mean survey magnitudes were $10.77\pm0.09$~mag,
$7.79\pm0.02$~mag, $0.17\pm0.15$~mag and $-2.03\pm0.01$~mag in 
{\it WISE} Bands W1 (3.3\mic), W2 (4.6\mic), W3 (12\mic) 
and W4 (22\mic) respectively;
the corresponding flux values were $15.13\pm1.19$~mJy (W1), 
$131.14\pm2.54$~mJy (W2), $24.84\pm3.52$~Jy (W3) and 
$53.63\pm0.05$~Jy (W4).

However the {\it WISE} survey data suffer from saturation at magnitudes 
brighter than 8.1 (W1), 6.7 (W2), 3.8 (W3) and 
$-0.4$ (W4); the survey magnitudes for Sakurai indicate that the {\it WISE}
photometry in bands W3 and W4 are affected by saturation,
with 30\% and 20\% of pixels, respectively, affected. Although data in the 
W1 and W2 bands are less affected by saturation, we do not
use data from the {\it WISE} cryogenic survey here; they are included in 
Table~\ref{spitzer} for completeness.

Although the {\it WISE} survey was completed in 2010 September with the exhaustion
of the cryogen, the mission was extended using bands W1 and W2 only to produce the
{\it NeoWISE} survey \citep{mainzer}. Sakurai was $\sim1$~mag brighter in both 
bands W1 and W2 during the {\it NeoWISE} survey 
($9.18\pm.0.02$~mag, $6.8\pm0.01$~mag respectively)
than it was $\sim6$~yr earlier, during the main {\it WISE} survey
(W1 and W2 magnitudes $10.77\pm0.09$~mag and $7.79\pm0.02$~mag respectively).
The {\it NeoWISE} data do not suffer from saturation, and 
we note from \cite{cutri} that no correction to the photometry is needed
in these data.
The averaged daily {\it NeoWISE} scans are shown in Fig.~\ref{K_lc}.

\subsection{\spitzer}

Sakurai was observed on several occasions  (Program IDs 3362, 30077, 40061; PI A.Evans)
in the period 2005--2008 with the Infrared Spectrometer \citep[IRS;][]{irs} on
the \spitzer\ \citep[][]{werner,gehrz}.
Spectra were obtained in both low- and high-resolution modes, covering 
the spectral range of 5-38\mic. 
In order to treat the data in a coherent and systematic way
we have used the data provided by the
Combined Atlas of Sources with Spitzer IRS spectra \citep[CASSIS;][]{cassis1,cassis2}.

\subsection{\SOFIA}
\label{sofia}
Spectra of Sakurai were obtained with \sof\ using the Faint Object infraRed CAmera 
for the \sof\ Telescope \citep[FORCAST;][]{forcast1,forcast2}.
The observations were carried out on two flights,
on 2014 March 25 (flight F155, Palmdale, California, Cycle~2 programme 
02\_0024, PI A.Evans) and
on 2016 July 11 (flight F318, Christchurch, New Zealand, Cycle~4 programme
04\_0003, PI A. Evans).
Sakurai was observed with the grisms, slits and
on-target integration times listed in Table~\ref{sof_log}. Observations
were generally carried out with the $4.7''$ slit; 
one observation on flight 
F318 used the $2.4''$ slit with the G063 grism.
All observations used the default 
two-position chopping with nodding mode.

\begin{table*}
\caption{Log of \sof\ FORCAST observations. \label{sof_log}}
 \begin{tabular}{ccccccc} 
  Flight & Date       & {Filter} &  Slit   & Wavelength & Nominal & Exposure \\
         & YYYY-MM-DD &          &  $('')$ &  coverage (\mic) & resolution $R$ & time (s) \\ \hline
  F155   & 2014-03-25 &  G111    &    4.7   & 8.4--13.7  & 130 & 1,500  \\
         &            &  G227    &    4.7   & 17.6--27.7 & 110 & 100 \\
         &            &  G329    &    4.7   & 28.7--37.1 & 160 & 100 \\
  F318   & 2016-07-11 &  G063    &    2.4   & 4.9--8.0   & 180 & 1,500  \\
         &            &  G111    &    4.7   & 8.4--13.7  & 130 & 1,500  \\
         &            &  G227    &    4.7   & 17.6--27.7 & 110 & 50    \\
         &            &  G329    &    4.7   & 28.7--37.1 & 160 & 500 \\\hline
 \end{tabular}
\end{table*}

The FORCAST scientific data products were retrieved from the \sof\ 
archive, after standard pipeline processing and flux calibration was 
performed \citep[for details see][]{Clarke2014}.
An extensive discussion of the FORCAST data pipeline can be found in the Guest 
Investigator Handbook for FORCAST Data 
Products, 
Rev.~B\footnote{{https://www.sofia.usra.edu/Science/DataProducts/FORCAST\\{\_}GI{\_}Handbook{\_}RevA1.pdf}}.

We note here that the slit widths in both the \spitzer\ and
\sof\ spectroscopic observations are such that any contribution from
the $40''$ PN (see Section~\ref{history}) is negligible.

\section{Dust emission}
\label{dustemission}
\subsection{The spectral energy distribution}
It is almost certain that the dust shell around Sakurai contains dust grains
that have a range of temperatures. Even so the observed SED of the dust resembles 
emission by a blackbody, particularly after 1998. 
We have therefore fitted
\begin{equation}
 f_\lambda = \frac{F}{\lambda^{5}} \: \frac{1}{\exp[B/\lambda]-1} 
 \label{bbfn}
\end{equation}
to the photometric and spectral data, where  
$T_{\rm dust}$ is the observed dust temperature,
$B~(=hc/{k}T_{\rm dust})$ and $F$ are constants to be determined. 
Obvious ``glitches'' in the data, possible and actual spectral features
\citep[such as HCN;][and see below]{evans-sak-sp}, as well as spectral 
regions where the quality of the data is clearly poor,
were removed before fitting. 

A sample of the fits for the de-reddened  UKIRT, \spitzer\ and \sof\ data is
shown in Fig.~\ref{sak_all}. Clearly in some cases the fit is poor at 
wavelengths $\ltsimeq2$\mic, reflecting
the fact that the emission in these cases is not well described by a simple 
blackbody. Indeed there is clear evidence in the earliest spectra for the 
presence of a photospheric contribution that diminishes with time.

The observed emitted power $f$ and dust luminosity $L_{\rm dust}$ were 
determined from (see Appendix~\ref{lfl})
\begin{eqnarray*}
[\lambda{f}_\lambda]_{\rm max} & = & F \left ( \frac{\alpha}{B} \right )^4 \, \frac{1}{e^\alpha - 1}  \\
f & = & 1.359 \: [\lambda{f}_\lambda]_{\rm max} \\
L_{\rm dust} & = & 4\pi{D}^2f \:,
\end{eqnarray*}
where $\alpha\simeq3.92$.
We use the fitted parameters $B$ and $T_{\rm dust}$ to determine 
$[\lambda{f}_\lambda]_{\rm max}$ as in most cases the wavelength at 
which $[\lambda{f}_\lambda]$ is a maximum falls 
outside the data wavelength range, or in a gap in the data. In doing so we risk 
missing some emission; however the SED of Sakurai, using near-contemporaneous 
UKIRT and JCMT data from late 2003 \citep{evans-sak}, suggests that the 1--5\mic\
and 450/850\mic\ emission lie on the same blackbody curve. Thus we are 
confident that we are not excluding any emission, for example by cold dust. 
The deduced values of $T_{\rm dust}$, $[\lambda{f}_\lambda]_{\rm max}$ and 
$L_{\rm dust}$ are given in Table~\ref{spitzer}. 

We note that the uncertainties in $T_{\rm dust}$ in Table~\ref{spitzer} are 
the formal uncertainties resulting from fitting Equation~(\ref{bbfn}) to
the data, and are not therefore ``physically'' meaningful. As already noted
the dust shell will include a range of temperatures that far exceed the
limits implied by the errors listed. However we use these errors to estimate
the uncertainties in the dust mass. Likewise the uncertainties in 
$L_{\rm dust}$ include the
uncertainties in both $[\lambda{f}_\lambda]_{\rm max}$ and in $D$. 
It is interesting to note that \cite{evans-sak} 
fitted a 360~K blackbody to the UKIRT/JCMT data by eye; the formal fit to the 
UKIRT data (Table~\ref{spitzer}) gives $359.6\pm0.2$~K.

Although there are only three photometric points in the 2MASS data, 
they all lie on the Wien tail of a blackbody and a fit to the data gives 
$T_{\rm dust}=950\pm37$~K, and 
[$\lambda{f}_\lambda$]$_{\rm max}=6.6[\pm2.5]\times10^{-12}$~W~m$^{-2}$.
The 2MASS spectral energy distributions (SED) is shown in 
Fig.~\ref{2mass-fig}, and the dependence of $T_{\rm dust}$, 
$[\lambda{f}_\lambda]_{\rm max}$ and $L_{\rm dust}$ 
on time is shown in Fig.~\ref{evol}.

\subsection{The dust temperature}

There is a monotonic decline in $T_{\rm dust}$, from $\sim1,200$~K to 
$\sim180$~K over the period 1998 August to 2016 July. We note that the 
observed dust temperature has been well below the formation temperature for amorphous 
carbon \citep[$\sim1,000$~K; see][and references therein]{gehrz18} since 1999.

For the simple case of dust moving away 
at constant speed from a source of bolometric luminosity $L_*(t)$, we 
would expect the dust temperature to vary as 
$T_{\rm dust}\propto({L_*(t)}/t^{2})^{1/(\beta+4)}$, where $\beta$ 
is the ``$\beta$-index'' for the dust material 
\citep[see below; and see][for derivations]{BE,DU}.
We therefore explore the possibility that the dust temperature declines 
with power-law dependence on time according to
\begin{equation}
 T_{\rm dust}= \frac{C}{(\mbox{MJD} - \mbox{MJD}_0)^\gamma} \:, \label{tbbd}
\end{equation}
where $C$, MJD$_0$ 
(the epoch at which the mass-loss leading to 
dust  condensation started) and $\gamma$ are constants to be determined; 
note that this expression implicitly assumes constant $L_*$. 

We have already noted that the 
drop in the optical brightness began as early as late 1997.
Fitting Equation~(\ref{tbbd}) to the data gives
\begin{eqnarray*}
 C & = & 3.38[\pm6.95]\times10^6\\
 \gamma & = & 1.14\pm0.23 \\
 \mbox{MJD}_0 & = & 49736 \pm  491 
\end{eqnarray*}
for $T_{\rm dust}$ in K (from Table~\ref{spitzer}) and MJD, MJD$_0$ in days.
The parameters MJD$_0$ and $\gamma$ are reasonably well constrained, while 
$C$ is not; this may be a reflection of the fact that $L_*$ is indeed 
time-dependent.  We revisit MJD$_0$ below.

\begin{figure*}
\setlength{\unitlength}{1mm}
\begin{center}
  \leavevmode
  \includegraphics[width=12cm,height=20cm]{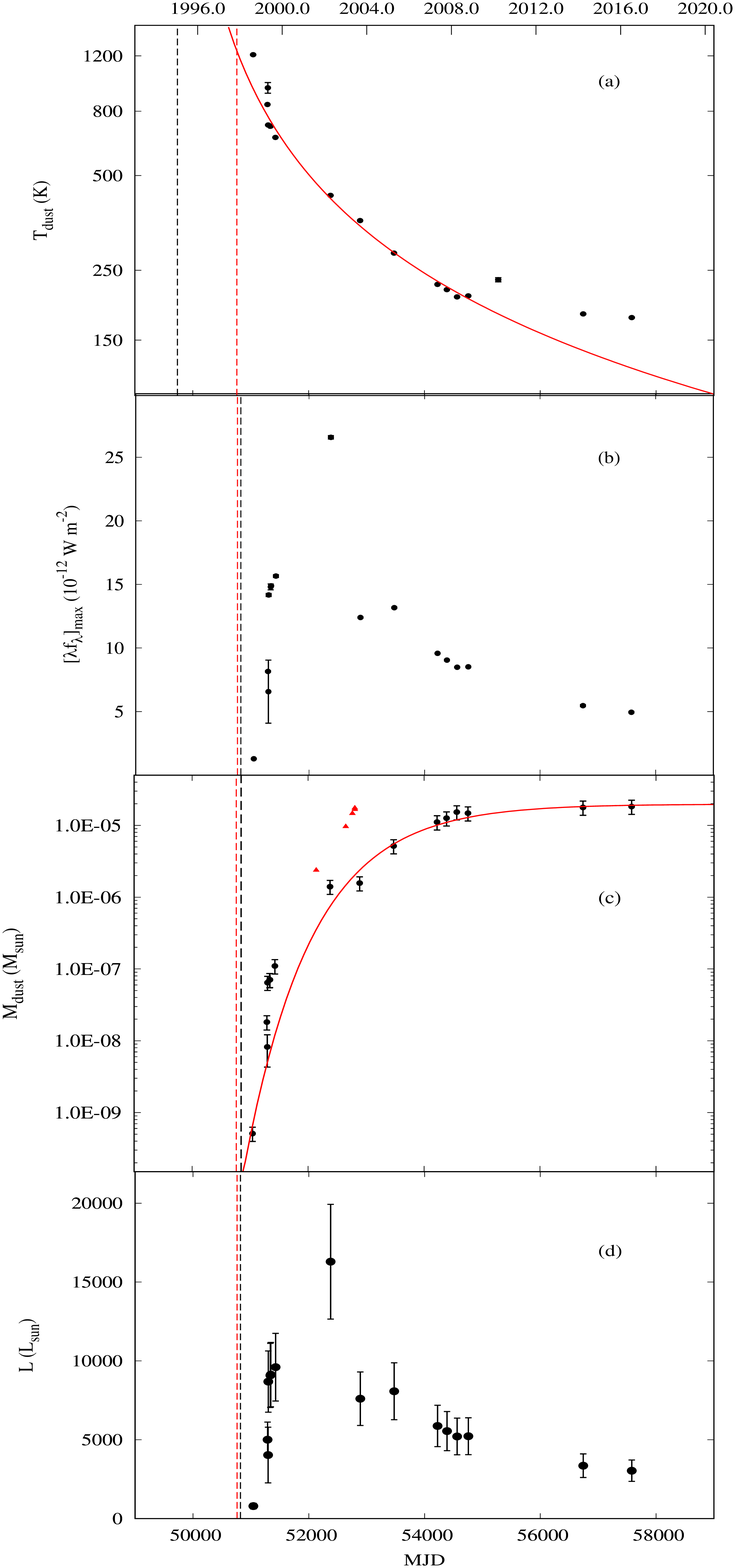}
  \caption[]{Evolution of Sakurai's dust shell.
  (a)~blackbody temperature of the dust;
  (b)~[$\lambda{f}_{\lambda}]_{\rm max}$;
  (c)~dust mass; red points are from \cite{evans-sak}, scaled to $D=3.8$~kpc;
  (d)~the bolometric luminosity of the dust. 
  Red curves are fits as described in text.
  Broken vertical black lines in (a) and (c) are epochs of dust formation as determined 
  separately for dust temperature and dust mass; broken vertical lines in (b)~and
  (d) are the weighted mean of epoch of dust formation as discussed in text.
  Broken vertical red lines are epoch of zero dust shell angular
  diameter, as determined by Evans et al. (2006). \label{evol}}
\end{center}
\end{figure*}

\subsection{The dust mass}
\label{Mdusts}
\cite{tyne-dusty} fitted {\sc dusty} \citep{dusty} models to UKIRT data covering
the period 1999 May 4 -- 2001 Sep 8. Assuming a gas-to-dust ratio of 200, 
they deduced a mass-loss rate that increased
from $\sim5\times10^{-6}$\Msun~yr$^{-1}$ at the earlier time to 
$\sim12\times10^{-6}$\Msun~yr$^{-1}$ at the later time.
However the gas-to-dust ratio in Sakurai's environment is
almost certainly very different from that assumed by \citeauthor{tyne-dusty}.

If we make the working assumption that the dust disc enveloping Sakurai is 
optically thin
at longer ($\gtsimeq5$\mic) wavelengths and at later times, then we can employ
the method used by \cite{gehrz-sgr} and \cite{evans-del} to estimate the dust mass, 
which is given by (see those papers for details)
\begin{eqnarray}
  \frac{M_{\rm dust}}{\Msun} & = & 4.81\times10^{16} \:
 \left [ \frac
 {[\lambda\,f_\lambda]_{\rm max}}
 {\mbox{W~m$^{-2}$}} \right ] \:
 \left [ \frac{D}{\mbox{kpc}} \right ]^2  \nonumber \\
 && \times \left [ \frac{\rho}{\mbox{kg~m$^{-3}$}} \right ] \:
 \frac{1}{A\,T^{(\beta+4)}} \:\:.
 \end{eqnarray}
Here $\rho$ is the density of the grain material \citep[1,500~kg~m$^{-3}$ for amorphous 
carbon, which we assume here;][]{jones}, and the constants $A$ and $\beta$ are defined such 
that the Planck mean absorption efficiency is given
by $\langle{Q}_{\rm abs}\rangle=AaT^\beta$, 
with $a$ being grain radius. For amorphous carbon, $A=58.16$ when $a$ is in cm, 
and $\beta=0.754$ \citep{evans-del}. Obviously the calculated dust masses 
can be scaled appropriately
if a different density is used for amorphous carbon. For a distance of 3.8~kpc, 
the dust mass is
\[ \frac{M_{\rm dust}}{\Msun} = 1.79\times10^{5} \left [ \frac
 {[\lambda\,f_\lambda]_{\rm max}}
 {\mbox{$10^{-12}$~W~m$^{-2}$}} \right ] \: T^{-(\beta+4)} \:\:. \]
We use the data in Table~\ref{spitzer} to find the dust masses in the final column. 
The uncertainties in $[\lambda\,f_\lambda]_{\rm max}$, $T$ and 
$D$ are propagated into the uncertainties in $M_{\rm dust}$;
however, there are likely to be substantial
uncertainties in both $A$ and $\rho$, which will make a (possibly) large
(and unquantifiable) contribution to the uncertainties 
in Table~\ref{spitzer}.

The dependence of $M_{\rm dust}$ with time 
is shown in Fig.~\ref{evol}. We see a clear and sustained increase in the dust mass 
from the time of formation in late 1997, until it levels off after MJD 
$\sim55000$ (mid-June 2009).
In Fig.~\ref{evol} we include the dust mass determined by \cite{evans-sak}
(see Section~\ref{nature}); although determined using rather different methods,
the dust masses deduced here and in \citeauthor{evans-sak} are reasonably
consistent.

Unlike the case for the dust temperature, there is no physically-based expression 
for the time-dependence of dust mass. For our present purpose, we fit a function of 
the form\footnote{The reason for this choice of function is that it results in a 
manageable time-dependence of the form: \\
$\log_{10}[{M_{\rm dust}]\propto}(1-\exp[-t/\tau_0])$.\label{fn}}
\begin{eqnarray}
 \frac{M_{\rm dust}}{10^{-10}\Msun} & = & 
      \exp  \left \{ K \, \ln{10} \times  \left (  1 - \exp   \left [ 
 -\frac{  \left   (  \mbox{MJD} - \mbox{MJD}_0  \right )} {\tau_0} 
 \right ]  \right ) \right \} 
 \nonumber  \\
 && 
  \label{Mdust}
\end{eqnarray}
to the variation of $M_{\rm dust}$ with MJD time, where $K$, $\tau_0$ and
MJD$_0$ (as before, the time of onset of 
mass loss) are constants to be 
determined\footnote{Eq.~(\ref{Mdust}) leads to
the somewhat spurious result that the dust mass $=10^{-10}$\Msun\ at MJD$_0$. 
This of course is a consequence of our choice of function, as described
in Footnote~\ref{fn}. This has no effect on our conclusions.}. 
We find that (with $M_{\rm dust}$ in \Msun\ and MJD in days)
\begin{eqnarray*}
 K & = & 5.30\pm0.16 \\ 
 \tau_0 & = & 1167\pm179 ~\mbox{days}\\
  \mbox{MJD}_0 & = & 50846\pm76
\end{eqnarray*}
give a passable fit to $M_{\rm dust}-t$ (see Fig.~\ref{evol}).
These values will depend somewhat on the form of the 
function assumed, but we might reasonably expect that the value of 
MJD$_0$ deduced here is consistent with that deduced from the time-dependence 
of the dust temperature ($49736\pm491$; see above); this they do, within the 
uncertainties. A weighted mean gives
\[ \mbox{MJD}_0 =50820\pm75 \:.\]
The mean value of MJD$_0$ corresponds to 1998 January 7, 
but the uncertainty allows the onset of 
the mass ejection that led to dust formation to have occurred
between MJD 50745 (1997 November 24) and MJD 50895 (1998 March 23).
This is close to the time at which the expansion of the dust shell 
started, on MJD$_0=50762$ \citep{evans-sak-sp,hinkle}.

As noted above, the values of $K$, $\tau_0$ and MJD$_0$ must depend on the
form of the function assumed for the fitting. Nevertheless, from
Equation~(\ref{Mdust}) we might expect that the dust mass approaches 
$10^{(K-10)}$\Msun\ as $\mbox{MJD}\rightarrow\infty$, i.e. 
$M_{\rm dust}(t\rightarrow\infty) \simeq 2.0\times10^{-5}$\Msun.
We also note that the dust mass on 2007 May 4 \citep[(MJD 54224.9, close to
the time of the VLTI observation of][MJD 54280]{chesneau}
was $\sim1.1\times10^{-5}$\Msun, somewhat less than the value given by
\citeauthor{chesneau}.

In principle, 
the time-derivative of $M_{\rm dust}$ in Eq.~(\ref{Mdust}) 
might give the mass-loss rate, in the form of dust, and scaling for an 
appropriate gas-to-dust ratio would give the overall mass-loss rate.
However given that (a)~the formula assumed is for mathematical convenience 
and has no physical basis (see footnote~\ref{fn}), 
(b)~the gas-to-dust ratio will likely change with time, and (c)~the deduced 
mass-loss in the form of dust alone might be due to grain growth rather than 
grain formation, this approach would not lead to a credible result. Indeed,
it is clear that the steep rise in the value of $M_{\rm dust}$ at the start 
of the dust phase results in a ferocious mass-loss rate, in dust alone. 
This leads us to suspect that the dust masses deduced -- at least at the earlier 
times -- do not accurately reflect the dust mass in Sakurai's environment. This is 
most likely because the dust emission is not as simple as we have supposed
here, and is unlikely to be optically thin at the shortest wavelengths
($\ltsimeq5$\mic), particularly after mid-1998 when the visual
light curve had entered the very   deep minimum. \cite{tyne-dusty}
concluded that the visual optical depth at these times is $\sim9$, leading to an 
optical depth $\sim2$ at 2.2\mic\ for any plausible extinction law. It is possible
that the dust mass is underestimated at the early times,
when the {\it UKIRT} observations were carried out.

We also note that determining the mass of circumstellar material,
or the mass-loss rate, from the dust parameters by assuming a specific
gas-to-dust ratio \citep[see e.g.][]{tyne-dusty} will almost
certainly lead to erroneous values 
because the gas-to-dust ratio is (a)~unknown
and (b)~almost certainly variable.

\subsection{The emitted power}

The power emitted by the dust  (see Fig.~\ref{evol})
seems to rise sharply after dust condensation began, reaching a maximum 
around MJD 51350, and then declining
(see Fig.~\ref{evol}). There appears to be a sharp peak in $L_{\rm dust}$ 
at around MJD~52378. However, as noted above (Section~\ref{Mdusts}), 
it is very likely that the dust is optically thick at the shorter IR wavelengths 
at the earliest times so that the $L_{\rm dust}$ at these times is underestimated.
This accentuates the sharpness of the peak in luminosity.

Although the dust shell is optically thick at visible wavelengths --
so that the dust shell could act as a ``calorimeter'' that monitors
the luminosity of the (invisible) central star -- we should be cautious of 
concluding anything about the behaviour of the central star from these data.
This is because interferometric observations \citep{chesneau} show that the
dust is distributed in a torus rather than completely enveloping the star,
so that an unknown 
(and almost certainly variable)
fraction of the star's radiation ``leaks out'' of the 
dust shell, along the axis of the torus and therefore away from our line 
of sight. The luminosity values in Table~\ref{spitzer} --
in which the errors include the uncertainties in 
$[\lambda{f}_{\lambda}]_{\rm max}$ and $D$ -- are therefore
lower limits on the stellar luminosity, $L_*$: $L_{\rm dust}=\phi{L}_*$,
where $\phi\le1$. 

However, if we suppose that the
same fraction of the star's light leaks out over the period 2005--2016,
then the relative changes in the $L_{\rm dust}$ values in Table~\ref{spitzer} 
might reflect the relative changes in the actual stellar luminosity.
The $L_{\rm dust}$ values in the table decline from $\sim16,000$\Lsun\ 
to $\sim3,000$\Lsun\ over an 11.25-year period. 
\cite{hajduk} have presented an evolutionary track for Sakurai. This suggests that the 
bolometric luminosity is expected to decrease from $\log[{L_*}/\Lsun]\sim4.2$ to
$\log[{L}_*/\Lsun]\sim 4.1$ over a similar period, compared with a decline
in $L_{\rm dust}$ from $\log[L_{\rm dust}/\Lsun]=4.21$ to
$\log[L_{\rm dust}/\Lsun]=3.48$: the observed decline is by a factor $\sim5.4$, 
compared to the predicted decline of $\sim1.25$.
However for reasons outlined above, our deduced dust luminosities
are almost certainly lower limits so comparison with stellar evolution models should not be taken too far.

\subsection{A fresh ejection event after 2008 October}
\label{fresh}

\begin{figure*}
\setlength{\unitlength}{1mm}
\begin{center}
  \leavevmode
    \includegraphics[width=8.5cm]{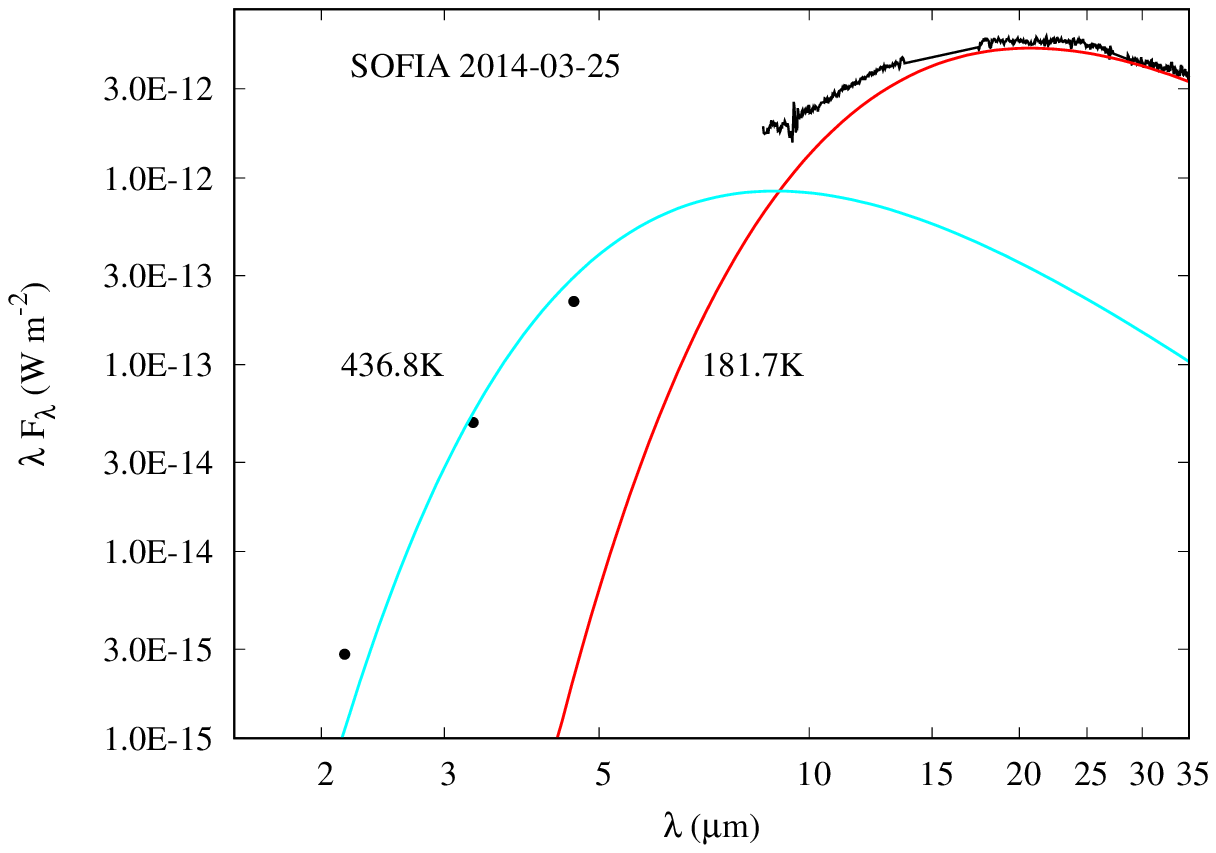}
        \includegraphics[width=8.5cm]{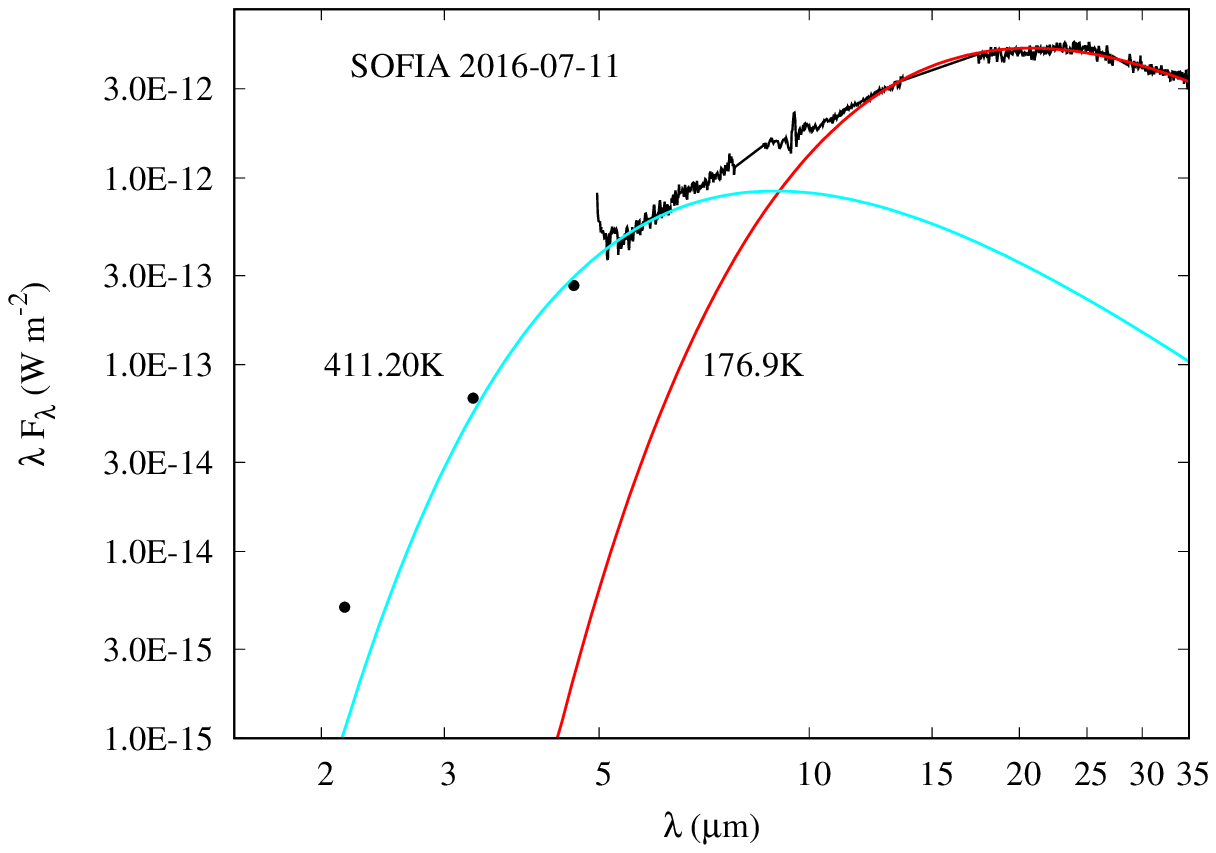}
  \caption[]{Evidence for formation of fresh dust.
  The \sof\ FORCAST data for both 2014 March 25 and 2016 July 11 show an excess 
  relative to the fitted blackbody at wavelengths $\ltsimeq10$\mic.
  Black points are near-contemporaneous $K_s$ and {\it NeoWISE} photometry. 
  Temperatures of the blackbodies, which are fitted separately to the long 
  wavelength data and the near-IR excess, are indicated.
  See text for details.
  The apparent ``feature'' just short of 10\mic\ in the \sof\ data is 
  the result of incomplete removal of telluric ozone.
  \label{LATE_SOFIA}}
\end{center}
\end{figure*}

Apart from a contribution from the stellar photosphere at the earliest 
times (see Fig.~\ref{sak_all}), the SED of the dust shell is generally 
well represented by a simple blackbody from 1999 to 2007. 
In the latest (\sof\ FORCAST) data, 
however, while the data for $\lambda\gtsimeq10$\mic\ are well described by 
a single blackbody, there appears to be an excess at the 
shorter ($\ltsimeq10$\mic)
wavelengths (see Fig.~\ref{LATE_SOFIA}; note that it is the properties 
of the cooler dust that are tabulated in Table~\ref{spitzer} and 
illustrated in Fig.~\ref{evol}). 
Furthermore, both sets of \sof\ data were obtained within days of the 
$K_s$ and {\it NeoWISE} data, and latter datasets confirm the excess 
seen by \sof\ (see Fig.~\ref{K_lc}).

We also note that,
in addition to the $\ltsimeq10$\mic\ excess, there might be a small $K_s$ band
excess in the latest SEDs (see Fig.~\ref{LATE_SOFIA}), 
so there might be a small stellar contribution to
the $K_s$ flux (see Section~\ref{Ks}). This suggests that the 
stellar photosphere might be finally beginning to peer through the gloom.
However the contribution can only be a minor one, and there can 
be no doubt that the great majority of this late $\ltsimeq10$\mic\ excess 
is due primarily to emission by dust.

Using the same procedure as in Section~\ref{Mdusts}, and again assuming amorphous
carbon, we determine the mass and temperature of the dust responsible for the 
near-IR excess. We find that the mass and temperature of the dust in 2014 March was
$2.6[\pm0.4]\times10^{-8}$\Msun\ and $437[\pm3]$~K respectively; in 2016 July 
these parameters had the values
$5.7[\pm1.0]\times10^{-8}$\Msun\ and $411[\pm6]$~K respectively.

The substantial rise in the $K_s$-band flux from  $\sim2009-2017$ (Fig.~\ref{K_lc}),
coupled with the rise in the {\it NeoWISE} Band W1 and W2 fluxes between 
2010 and 2016 (Section~\ref{wise} and Fig.~\ref{K_lc}), the apparent increase in 
dust mass and the cooling of the ``new'' dust between 2010 and 2014-16, 
all point to renewed, and possibly substantial, mass-loss since $\sim2008$. 
The constancy of the molecular column densities from 2005 April -- 2008 October (see
Section~\ref{mabs} below) suggests that the onset of this most recent mass-loss 
episode must have occurred  after 2008 October.

The mass of the fresh 
dust is some three orders of magnitude smaller than that of the dust formed in the first
mass-loss episode (cf. Table~\ref{spitzer}). However if the fresh dust-formation episode
follows the same pattern as the earlier event, it does not augur well for our
prospects of seeing the stellar photosphere any time soon.

\section{Possible dust features}

\begin{figure*}
\setlength{\unitlength}{1mm}
\begin{center}
  \leavevmode
    \includegraphics[width=8cm]{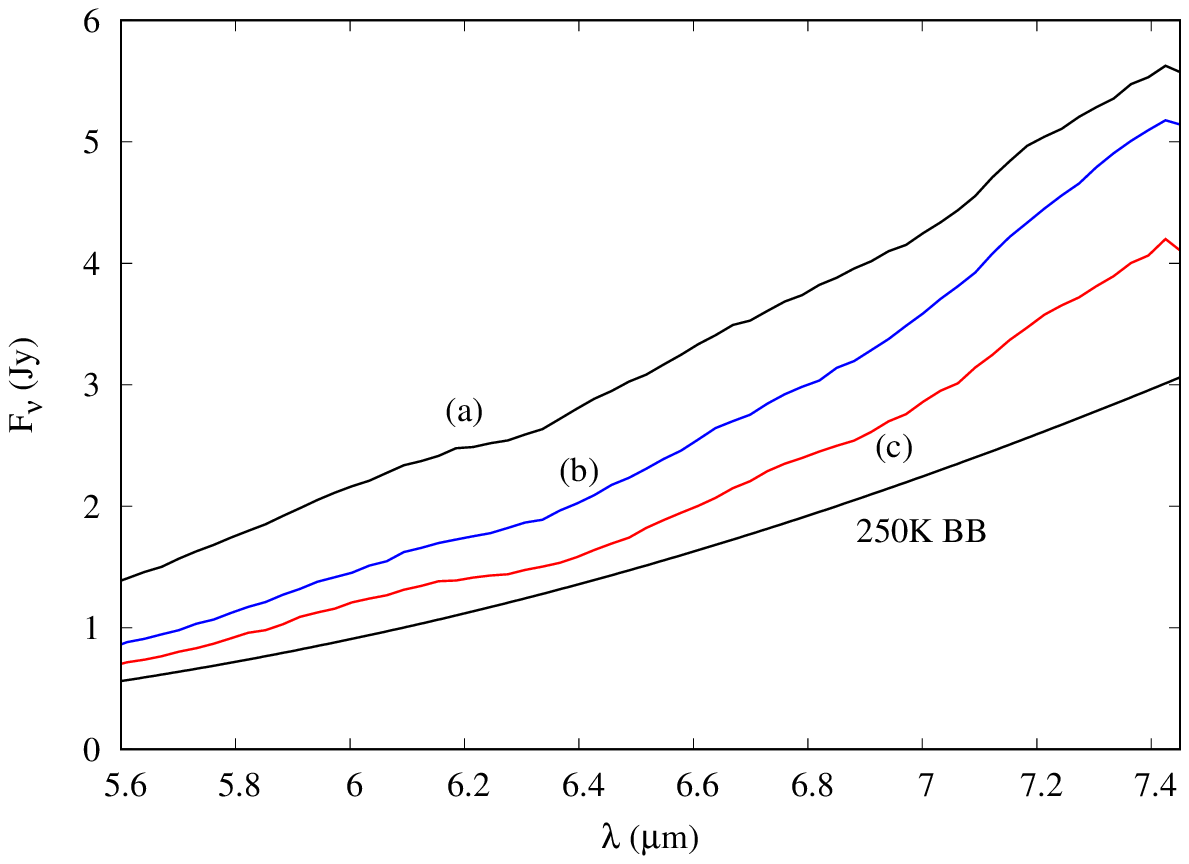}
  \includegraphics[width=8cm]{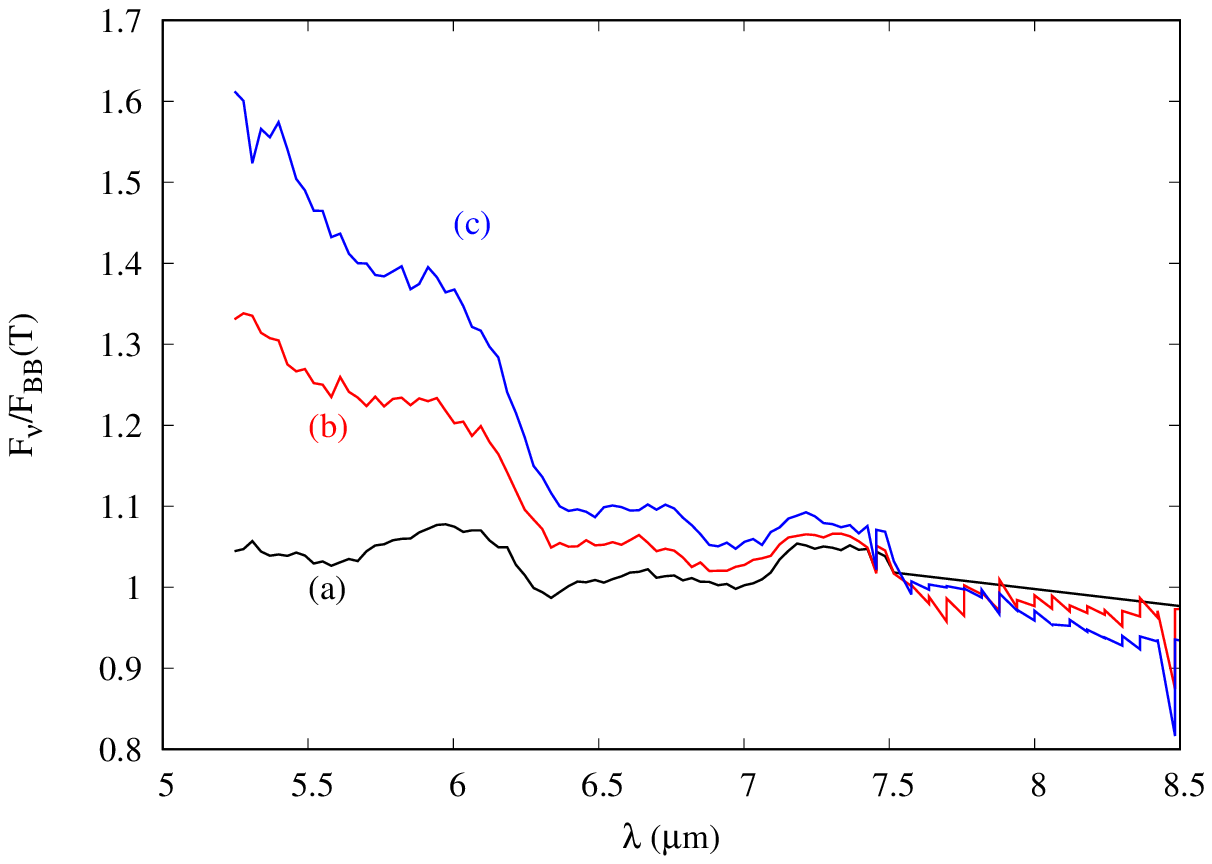}
  \caption[]{Left: \sirtf\ spectra of Sakurai in the region 5.6--7.4\mic\ for 
  (a) 2005-04-15 (283.5~K), (b) 2007-05-04 (225.5~K), (c) 2007-10-15 (217.0~K). 
  A 250~K blackbody, arbitrarily normalised, is shown for comparison.
  Right: Data from left panel, rectified by fitted blackbodies as discussed in text, 
  to bring out the weak features; labelling of curves is as per the left panel.
  \label{6mic}}
\end{center}
\end{figure*}

We consider here the possibility that there might be dust features, either in 
emission or in absorption in the mid-IR spectra of Sakurai.

In Fig.~\ref{6mic} (left panel) we show \sirtf\ IRS spectra for three epochs, 
together with a 250~K blackbody curve for comparison. The latter is obviously 
smooth, whereas the \sirtf\ spectra are not. There are weak inflections in 
the data that hint at the presence of features in 
either absorption or emission. The putative features are too broad to be 
atomic/ionic, and do not have the profiles expected of molecular features.
Also in Fig.~\ref{6mic} (right panel) the same data are displayed as 
$f_\nu/f_{\rm BB}(T)$, where $T$ is the temperature of the best fitting 
blackbody from Table~\ref{spitzer}, to highlight these features 
(unfortunately the \sof\ data are too noisy in this 
spectral region to confirm the presence of the features).

While it might be possible to choose a better continuum to bring out the features, 
it is clear that the overall spectral shape is the same for all three epochs in 
Fig.~\ref{6mic}. We consider that they are weak absorption features, with maximum 
absorption at $\sim6.3$\mic\ and $\gtsimeq7.5$\mic, and a weaker feature longward of 
$\sim7$\mic. A plausible identification is with hydrogenated amorphous carbon (HAC), 
and specifically nitrogenated HAC (see below).

\cite*{garcia} have identified similar (weak) features in \sirtf\ IRS spectra 
of RCB stars. They find that these features are somewhat different in the 
``hydrogen-rich'' RCB stars and in the more common ``hydrogen deficient'' RCB stars. 
The former have features at $\sim6.27$\mic, $\sim6.6$\mic, $\sim7.02$\mic\ and 
$\sim7.7$\mic, rather similar to those we identify in Sakurai.

Laser vaporisation of graphite in a H$_2$/N$_2$ gas mixture gives rise to
features in the 6.0--6.5\mic\ range and a weaker feature at $\sim7.0$\mic,
together with broad absorption going from $\sim7.5$\mic\ to longer wavelengths
\citep{grishko}, just as we see in Sakurai. As we discuss in Section~\ref{mabs} 
below, we see HCN, HNC and 
other small hydrocarbon molecules in absorption against the dust shell. 
Given that there have been at least two dust ejection
events in the past $\sim20$~years (see Section~\ref{fresh}), the 
absorbing material may arise in material from an even earlier ejection  event.
The possible nitrogenation of the HAC also seems consistent with the presence
of N-bearing molecules.

The observational \citep{garcia} and laboratory evidence \citep{grishko} seem
therefore to point to the existence of a cooler, hydrogen-rich carbon dust shell
outside the dust shell that is prominent in the near- and mid-IR.

\begin{figure}
\setlength{\unitlength}{1mm}
\begin{center}
  \leavevmode
 \includegraphics[width=8.cm]{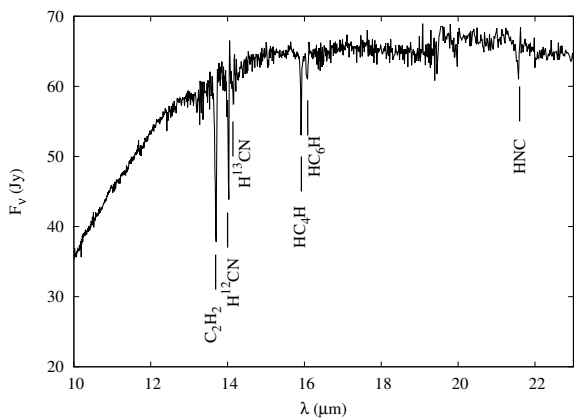}
 \caption[]{\sirtf\ spectrum showing absorption by small hydrocarbon molecules 
 in Sakurai on 2005 April 15. Data de-reddened as described in text.  
  \label{AF}}
\end{center}
\end{figure}

\section{Molecular absorption}
\label{mabs}

\begin{figure}
\setlength{\unitlength}{1mm}
\begin{center}
  \leavevmode
\includegraphics[width=8cm]{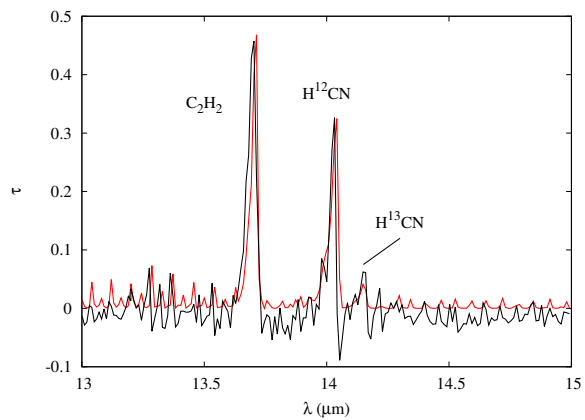}
\caption[]{Fit of slab model as described in the text to H$^{12}$CN, 
H$^{13}$CN and C$_2$H$_2$ for the 2005 April 15 \sirtf\ spectrum. 
Black curve, optical depth derived as described in text; red curve, 
fit with parameters in Table~\ref{column}. 
\label{FIT}}
\end{center}
\end{figure}

\begin{figure}
\setlength{\unitlength}{1mm}
\begin{center}
  \leavevmode
 \includegraphics[width=8.cm]{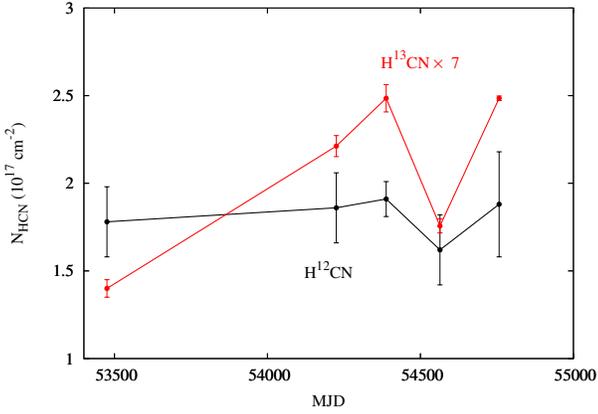}
  \caption[]{Variation of the HCN column densities with time as determined 
  from the 14\mic\ feature in the \sirtf\ spectra. 
    The H$^{13}$CN data are multiplied by 7.
  \label{HCN}}
\end{center}
\end{figure}

\begin{table}
\caption{Hydrocarbon and other small molecules in Sakurai as observed by \sirtf. 
H$^{12}$CN transitions from Aoki et al. (1999); HNC transitions from Harris 
et al. (2002). Tentative identifications are indicated by a ``[?]''.
\label{hcmols}}
\begin{tabular}{lccc}
           &                            &                       &                          \\      
Molecule   & $\lambda_{\rm obs}$ (\mic) & Band                  & $\lambda_{\rm id}$ (\mic)  \\ \hline 

C$_2$H$_2$ & 13.69                      & $\nu^5$ Q branch      & --       \\
H$^{12}$CN & 14.03                      & $2\nu^2_2-1\nu^1_2$   & 14.00      \\
H$^{12}$CN &                            & $1\nu^1_2-0\nu^0_2$   & 14.04       \\
H$^{13}$CN & 14.14                      &   Q                   & 14.16    \\
HC$_4$H    & 15.92                      & $\nu_8$               & 15.93    \\
HC$_6$H    & 16.09                      & $\nu_8$               & 16.07     \\
  ~~~$''$  &                            & $\nu_{11}$            & 16.10     \\
HNC        & 21.61                      &                       & 21.61   \\   
CH$_3$CCH [?] &    30.51                    &                       & 30.488   \\
C$_2$N  [?]   &   30.88                    &                       & 30.864    \\ \hline
       \end{tabular}
\end{table}

\cite{evans-sak-sp} reported the detection of a number of small hydrocarbon molecules 
in the \sirtf\ spectrum of Sakurai; these are listed in Table~\ref{hcmols} and illustrated 
in Fig.~\ref{AF}. To the list in \citeauthor{evans-sak-sp} we have added a number of new 
identifications, which are further discussed below.

\begin{table*}
\caption{Column densities of HCN isotopologues and  C$_2$H$_2$. Assumed gas temperature 
$T=400$~K.\label{column}}
\begin{tabular}{lccccc}
Date       &   MJD    &  \multicolumn{3}{c}{Column density}  & $^{12}$C/$^{13}$C$^\dag$\\ \cline{3-5} 
&&&&&\\
           &          &       H$^{12}$CN     &  H$^{13}$CN & C$_2$H$_2$ & \\ \hline 
           &          &   \multicolumn{2}{c}{($10^{17}$ cm$^{-2}$)} & ($10^{18}$ cm$^{-2}$) & \\ \hline
 2005-04-15 & 53475.3  &  $1.78^{+0.22}_{-0.19}$ & $0.20^{+0.05}_{-0.04}$   & $1.15^{+0.30}_{-0.19}$ & $8.9\pm0.3$  \\ 
 &&&&&\\
2007-05-04 & 54224.9  &  $1.86^{+0.13}_{-0.20}$ & $0.31^{+0.06}_{-0.04}$   & --- & $5.9\pm0.5$  \\
&&&&&\\
2007-10-15 & 54388    &  $1.91^{+0.23}_{-0.28}$ & $0.35^{+0.15}_{-0.10}$   & --- & $5.5\pm0.7$   \\
&&&&&\\
2008-04-08 & 54564    &  $1.62^{+0.20}_{-0.21}$ & $0.25^{+0.04}_{-0.04}$   &$0.91^{+0.18}_{-0.19}$ & $6.5\pm0.4$   \\
&&&&&\\
2008-10-18 & 54757.1  &  $1.88^{+0.30}_{-0.30}$ & $0.36^{+0.07}_{-0.13}$   & --- & $5.2\pm0.6$ \\ \hline
\multicolumn{6}{l}{$^\dag$From the HCN isotopologues.}\\
       \end{tabular}
\end{table*}

\begin{figure}
\setlength{\unitlength}{1mm}
\begin{center}
  \leavevmode
\includegraphics[width=7.5cm]{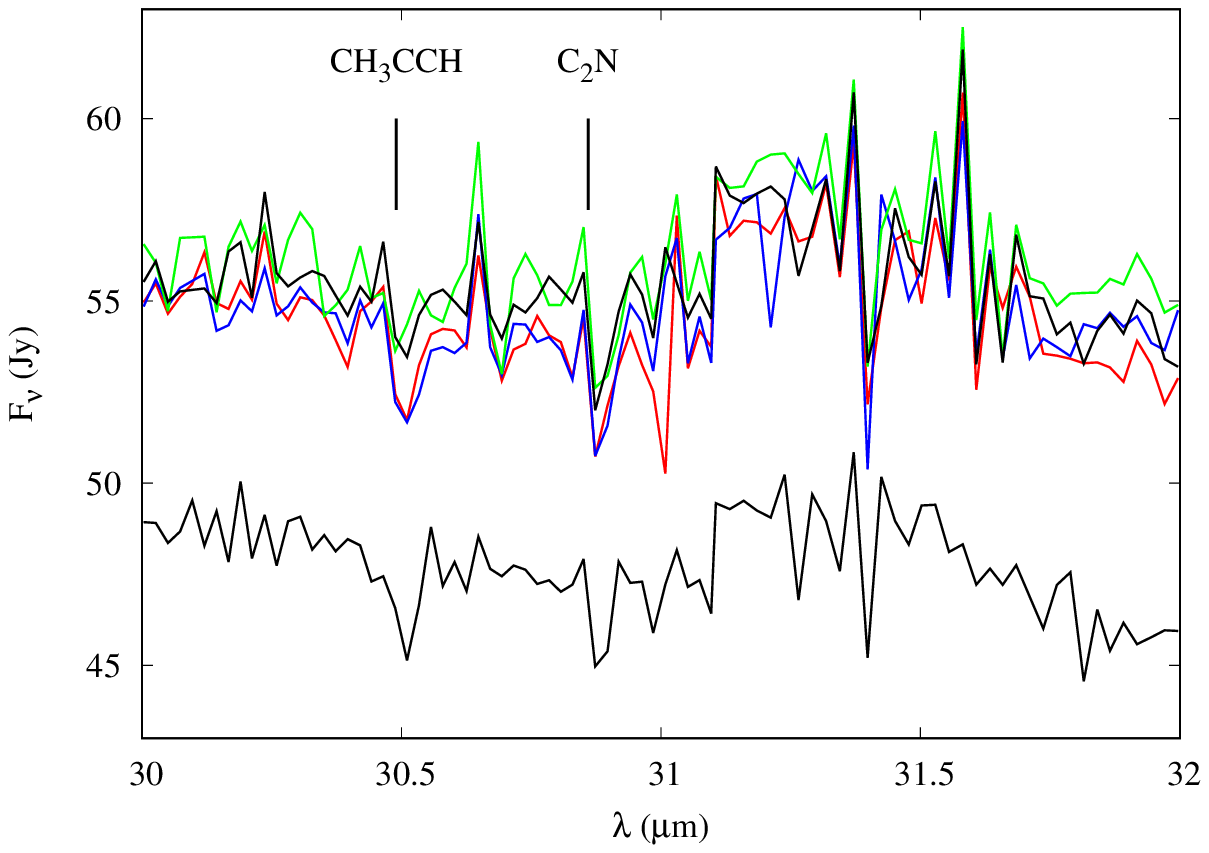}
    \caption[]{Absorption by CH$_3$CCH and C$_2$N in the spectral 
  range 30--32\mic. Upper black: 2005-04-15; red: 2007-05-04; blue: 2007-10-15; 
  green 2008-04-08; lower black: 2008-10-18. 
    Data de-reddened as described in text.\label{benz}}
\end{center}
\end{figure}

In \cite{evans-sak-sp} we used a simple model to estimate the column densities and 
temperatures of the H$^{12}$CN, H$^{13}$CN and C$_2$H$_2$ molecules, and we reprise 
this exercise here. We determine the continuum in the region of each feature by 
fitting a linear function $F_\nu \mbox{(Jy)} = a_0 + a_1\lambda(\mic)$, whence the 
wavelength-dependence of the optical depth $\tau$ is obtained. The typical 
uncertainty in $\tau$ is $\pm0.02$, as determined by the rms value of the ``zero'' 
$\tau$ in the 13--15\mic\ range. The optical depth is fitted assuming that the 
absorbing gas is homogeneous and isothermal, in thermodynamic equilibrium, and 
located in a plane-parallel slab in front of the dust. Individual transitions are 
assumed to have intrinsic widths determined by the velocity widths of the HCN 
isotopologues measured by \cite{tafoya} ($\simeq300$\vunit).

\subsection{HCN}
\label{hcn}
\subsubsection{The $^{12}$C/$^{13}$C ratio}
Using the strong H$^{12}$CN feature at 14.03\mic\ in the \sirtf\ data, we 
find that the temperature is not particularly well constrained, with values 
lying between 350~K and 450~K; this is a little lower than that determined 
by \citeauthor{evans-sak-sp} (2006, $\sim400-500$~K). We attribute this difference to 
two factors: (a)~the greater line width assumed here ($\simeq10^{-3}\lambda$) compared 
to our previous work -- leading to a lower temperature, and (b)~the fact 
that, in the present work, we 
use spectra extracted in a uniform way from the CASSIS archive, whereas in 
\citeauthor{evans-sak-sp} the data were reduced ``manually'' using the  
\cite{spice} software package. 

\cite{tafoya} detected the $J=4\rightarrow3$ and $1\rightarrow0$ HCN rotational 
transitions in Sakurai, and deduced a rotational temperature $13\pm1$~K. They attributed 
the difference between their value and that deduced by \cite{evans-sak-sp} to the 
fact that the gas they detected is observed over the entire circumstellar region, 
including cooler, more diffuse gas, whereas the \sirtf\ observations sampled hotter, 
denser, gas close to the dust shell. Further, the deduced gas temperature is greater 
than the corresponding temperature of the dust against which the features are seen 
in absorption. Clearly the properties of the dust and gas in Sakurai's environment
are far more complex than we have assumed here.

For our present purposes we take a temperature
of 400~K for the molecular gas, and
the fit for 2005 April 15 is shown in Fig.~\ref{FIT}.
For each date, we have determined the corresponding column density for 
both H$^{12}$CN and H$^{13}$CN (see Table~\ref{column}).
The variation of the absorbing column density with time, for both isotopologues, is shown in Fig.~\ref{HCN}. 

For the simple spherically-symmetric
case, the column density $N$ is related to the mass-loss rate $\dot{M}$ by 
\[ N = \frac{\dot{M}}{4\pi{v}\mu{R_1}} \left (1 - \frac{R_1}{vt} \right)\]
where $v$ is the wind speed,
$\mu$ is the mean molecular mass and $R_1$ is the radius at the base of the wind.
Taking the stellar luminosity $L_*$ and temperature $T_*$ from \cite{hajduk} 
to estimate the stellar radius $R_*$, and assuming $R_1\simeq{R_*}$, we get $R_1\simeq2.3\times10^9$~m. 
With a wind speed of 300\vunit\ \citep{tafoya} for $t\sim20$~yr, the 
factor in brackets is close to unity, so that the column density is simply proportional to $\dot{M}$. There is no compelling evidence for any 
variation of the column density with time, suggesting that the mass-loss rate 
as measured by HCN was $\sim$~constant 
over the period 2005 April to 2008 October. 
The mean column density in H$^{12}$CN is 
$1.8\times10^{17}$~cm$^{-2}$, a little larger than that found by 
\cite{evans-sak-sp}.

The ratio of the H$^{12}$CN and H$^{13}$CN column densities gives an estimate of the
$^{12}$C/$^{13}$C ratio.
It is evident from Fig.~\ref{HCN} that -- as reported by numerous authors
\citep{pavlenko04,evans-sak-sp,worters,tafoya} -- the $^{12}$C/$^{13}$C ratio is 
very low. A mean of the values in Table~\ref{column} gives  
$^{12}\mbox{C}/^{13}\mbox{C}=6.4\pm0.7$, about $1/14$ of the solar value.

\subsubsection{Implications}

If the high abundance of $^{13}$C derived above is a consequence of proton capture,
the amount of $^{13}$C produced depends on the amount of hydrogen initially present 
in the envelope. Effectively each $p$ gives rise to a $^{13}$C,
so the production of a mass of $^{13}$C, $M_{13}$, requires the processing of 
$\sim{M}_{13}/13$ of hydrogen.
Assuming that the relative numbers of $^{12}$C and $^{13}$C in the dust are 
the same as those in the gas phase, the dust mass of 
$2.0\times10^{-5}$\Msun\ implies a mass of $^{13}$C in the dust of $\sim{M}_{\rm dust}/6.9\simeq2.9\times10^{-6}$\Msun.
This requires the processing of $\sim2.2\times10^{-7}$\Msun\ of hydrogen, and
an envelope mass of $\sim3\times10^{-7}$\Msun\ assuming solar abundances in the envelope.

As already noted (Section~\ref{history}) we can rule out a progenitor with mass less 
than 1.25\Msun\ for Sakurai. The envelope mass before the VLTP is given as a function of
progenitor mass by \cite{MB}; his Figure~7 implies that the envelope on 
Sakurai's progenitor had mass at most $\sim1.5\times10^{-4}$\Msun, which would
likely have resulted in $\sim2\times10^{-3}$\Msun\ of $^{13}$C. While some of
the $^{13}$C would have been further processed, the 
mass of $^{13}$C alone seems consistent with the mass of carbon dust we have deduced.

\subsection{C$_2$H$_2$}

The isotopologues of acetylene display numerous ro-vibrational bands around 13\mic\ 
\citep[see][especially their Figure~3, for details]{chernicharo}. We see these in
absorption in two \sirtf\ IRS spectra, namely those obtained on 2005 April 15 
\citep[discussed in][]{evans-sak-sp} and 2008 April 8 (see Fig.~\ref{AF}); the 
\sirtf\ IRS spectra obtained on other dates are not of sufficient quality around 
13\mic\  to reveal these features. 

We have modelled these, for both $^{12}$C$_2$H$_2$ and $^{13}$C$^{12}$CH$_2$, using 
molecular data from the High Resolution Transmission molecular absorption database
\citep[HITRAN;][]{gamache,rothman}; the two isotopologues are not resolved in our 
data. We again assume a gas temperature of 400~K. We also assume that 
the relative abundance of the two isotopologues is determined simply 
by the $^{12}$C/$^{13}$C 
ratio (6.4) determined from the HCN data, although this is likely 
to be an over-simplification.

The resulting column densities for C$_2$H$_2$ are given in 
Table~\ref{column}, and the fit for 2005 April 15 is included in 
Fig.~\ref{FIT}. There is no evidence that the C$_2$H$_2$ column density varied 
between 2005 April and 2008 October, and the mean column density  is 
$1.0[\pm0.3]\times10^{18}$~cm$^{-2}$.
This value is similar to that given in \cite{evans-sak-sp}.

\subsection{Other features}

In addition to HCN (Section~\ref{hcn}) we report here the secure identification
of the fundamental
transition of hydrogen isocyanide (HNC) at 21.61\mic\ \citep*{harris}.
HNC is an isomer of HCN but lies high in energy ($\sim3,700$~eV) above HCN.  
The high abundance of HNC in interstellar clouds is generally attributed to the 
electron recombination  with HCNH$^+$ \citep{graninger}.
Its presence in Sakurai could be an indication of ion-molecule chemistry.

In addition to the above molecular features, there are  two 
weak features at 30.51\mic\ and 30.88\mic, which we tentatively identify 
with methyl acetylene (aka propyne; CH$_3$\chemone{C}\chemthree{CH}) 
and  the radical C$_2$N respectively.  If correct, we believe that this is the 
first astrophysical observation of methyl acetylene in the IR.
It has been observed in {\em young} ($\ltsimeq10^3$~yr) 
PNe at millimetre wavelengths
\citep[see e.g.][and references therein]{schmidt}, suggesting an origin
in the Sakurai environment resulting from a recent mass-loss episode, 
such as that evidenced by the excess in Fig.~\ref{LATE_SOFIA}.
All subsequent \sirtf\ spectra display these features, which give 
us confidence in their reality; other apparent features either are present in only 
one spectrum, are present in only one spectral element, or are the result of mismatch 
in the flux scale between \sirtf\ grisms.
However the features in Table~\ref{hcmols} and Fig.~\ref{AF} do not appear in the 
\sof\ spectra, either because they have weakened to the extent that they are no 
longer detectable, or because they fall in spectral gaps between wavelength ranges 
covered by the \sof\ FORCAST grisms.

Methyl acetylene -- in which one of the H atoms in acetylene is replaced 
by CH$_3$ -- 
is commonly seen in dense interstellar molecular clouds, for which it is sometimes 
used as a thermometer \citep[e.g][]{askne,kuiper}. \cite*{hickson} have argued 
that the formation of methyl acetylene is problematic in the gas phase (at least 
in molecular cloud and interstellar medium environments), and that it may most 
easily form via surface hydrogenation of C$_3$ on grain surfaces. Such a process
must have occurred in the hydrogen-rich material ejected by Sakurai before its
VLTP phase.

The formation of  C$_2$N has been discussed -- in the context of both interstellar 
and circumstellar environmemts -- by \cite{mebel}, who suggest it might be formed 
via the route
\[ \mbox{C} + \mbox{HCN} \longrightarrow \mbox{C}_2\mbox{N} + \mbox{H} \:\:.\]
\citeauthor{mebel} note that this reaction is unlikely to proceed in the low 
temperatures in interstellar space but is likely to occur in warmer circumstellar 
environments. In view of the presence of HCN in Sakurai's environment, the 
presence of C$_2$N is not unexpected
\citep[although it was not detected in the archetypical carbon star 
IRC+10216 by][]{fuchs}.

\section{Conclusions}

We have presented a 20-year overview of the IR development of V4334~Sgr (Sakurai's
Object), highlighting the properties of the dust, and the evolution of the 
molecular absorption. Our principal findings and conclusions are as follows:
\begin{enumerate}
\itemsep=0mm
 \item the observable dust around Sakurai has cooled from $\sim1,200$~K in
1998 to $\sim180$~K in 2016, and some $2\times10^{-5}$\Msun\ of amorphous
carbon dust has formed between the commencement of the first dust phase 
in 1997 and the last \sof\ observation in 2016; 
\item there is evidence, in the form of dust that is hotter than that ejected
between 1998 and 2008, for a more recent dust ejection event, which
commenced after 2008 October;
\item we have identified a number of small hydrocarbon and other molecules 
in absorption against the dust shell; 
\item we find no significant variation in the column densities of
HCN and C$_2$H$_2$ between 2005 April and 2008 October;
\item we find possible evidence for absorption by nitrogenated HAC;
\item the $^{12}$C/$^{13}$C ratio, based on the column densities of the 
HCN isotopologues, is found to be $6.4\pm0.7$, which 
is 14 times smaller than the solar value, and consistent with Sakurai 
being a VLTP object.
\end{enumerate}

\section*{Acknowledgments}

We thank the referee for their careful and thorough
review of the first version of this paper, and for their constructive
and helpful suggestions.

Based in part on observations made with the NASA/DLR Stratospheric Observatory 
for Infrared Astronomy (SOFIA). 
SOFIA is jointly operated by the Universities Space Research Association, Inc. 
(USRA), under NASA contract NNA17BF53C, and the Deutsches SOFIA Institut (DSI) 
under DLR contract 50 OK 0901 to the University of Stuttgart.

This publication makes use of data products from the Wide-field Infrared Survey 
Explorer ({\it WISE}), which is a 
joint project of the University of California, Los Angeles, and the Jet Propulsion 
Laboratory/California Institute of Technology, funded by the National Aeronautics 
and Space Administration.

It also makes use of data products from the Two Micron All Sky Survey, which is 
a joint project of the University of Massachusetts and the Infrared Processing and 
Analysis Center/California Institute of Technology, funded by the National Aeronautics 
and Space Administration and the National Science Foundation.

Based on data products from VVV Survey observations made with the VISTA 
telescope at the ESO Paranal Observatory under programmes IDs 179.B-2002 
and 198.B-2004. This research has made use of the services of the ESO 
Science Archive Facility.

RDG was supported by NASA and the United States Air Force.
CEW acknowledges support from  \sof\ and NASA.
DPKB is supported by a CSIR Emeritus Scientist grant-in-aid which is 
being hosted by the Physical Research Laboratory, Ahmedabad.
TRG's research is supported by the Gemini Observatory, which is operated by the Association of Universities for Research in Astronomy, Inc., under a cooperative agreement with the NSF on behalf of the Gemini partnership: the National Science Foundation (United States), National Research Council (Canada), CONICYT (Chile), Ministerio de Ciencia, Tecnolog\'ia e Innovaci\'on Productiva (Argentina), Minist\'erio da Ci\^encia, Tecnologia e Inova\c c\~ao (Brazil), and Korea Astronomy and Space Science Institute (Republic of Korea).
PJS thanks the Leverhulme Trust for the award of a Leverhulme Emeritus Fellowship.
SS acknowledges partial support from NASA and HST grants to ASU.
TL acknowledges financial support from GA\v{C}R (grant number 17-02337S).
The Astronomical Institute Ond\v{r}ejov is supported by the project
RVO:67985815. This project has received funding from the European Union’s
Framework Programme for Research and Innovation Horizon 2020 (2014-2020)
under the Marie Sk\l{}odowska-Curie Grant Agreement No. 823734.

\appendix

\section{The observed flux}
\label{lfl}
For a blackbody emitter having  temperature $T$, and flux density
\[ f_\lambda = \frac{F}{\lambda^5} \:\: \frac{1}{\exp\left(B/\lambda\right)-1} \:,\]
where $F$ and $B=hc/kT$ are constants,
it is trivially shown that the wavelength $\lambda_{\rm m}$  at which $[\lambda{f}_\lambda]$ is a maximum
is $\lambda_{\rm m}=B/\alpha$; here $\alpha$ ($\simeq3.9207$\ldots) is the solution of the transcendental equation
\[  4\,(1 - e^{-\alpha}) = \alpha . \]
Thus 
\[ [\lambda{f}_\lambda]_{\rm max} = F \left ( \frac{\alpha}{B} \right )^4 \frac{1}{e^\alpha-1} \:,\] 
as used in Section~\ref{dustemission}.

Furthermore, the total observed flux
$f$ for a blackbody source of radius $R$ and distance $D$, in the absence of an intervening medium, is
\begin{equation}
f = \int_0^\infty f_\lambda d\lambda = \frac{\pi R^2}{D^2} \frac{\sigma
T^4}{\pi} \:.
\label{f1}
\end{equation}
For a blackbody, 
$f$ is related to $[\lambda f_\lambda]_{\rm max}$ by:
\begin{equation}
f = \frac{\pi^4}{15} [\lambda f_\lambda]_{\rm max} \left (
\frac{e^\alpha-1}{\alpha^4} \right ) \simeq 1.359\ldots[\lambda f_\lambda]_{\rm
max} \: ,
\label{f2}
\end{equation}
where $\sigma$ is the Stefan-Boltzmann constant, as used in Section~\ref{dustemission}.

Finally, the ``blackbody angular diameter'' $\theta_{\rm BB}$  is given by 
\[
\theta_{\rm BB} =  \frac{2R}{D} = \frac{2.331}{\sigma^{1/2}T^2} \: [\lambda
f_\lambda]_{\rm max}^{1/2} \]
so that 
\[ \theta_{\rm BB} = 2.02\times10^{9}\:[\lambda  f_\lambda]_{\rm max}^{1/2}\:T^{-2} \: , \]
where $\theta_{\rm BB}$ is in arcsec and
$[\lambda f_\lambda]_{\rm max}$ is in W~m$^{-2}$.

\bsp

\label{lastpage}

\end{document}